\documentclass[final,5p,times,twocolumn]{elsarticle}

\usepackage{balance}
\usepackage[scr=boondox]{mathalfa}
\usepackage{pdflscape}
\usepackage[nocomma]{optidef} % For nice optimization models: https://mirror.math.princeton.edu/pub/CTAN/macros/latex/contrib/optidef/optidef.pdf
\usepackage{lipsum}
\usepackage[normalem]{ulem}
\usepackage[utf8]{inputenc}
\usepackage{multirow, multicol, makecell, booktabs, graphicx}
\usepackage{afterpage}
\usepackage{placeins}     % For \FloatBarrier
\usepackage{dsfont}
\usepackage{adjustbox}
\usepackage{url}
\usepackage[utf8]{inputenc}
\usepackage{amsmath,bm}
\usepackage{hyperref} %Hyperreferences
\hypersetup{
     colorlinks   = true,
     citecolor    = blue,
     linkcolor = blue,
     urlcolor = blue
}
\usepackage[switch]{lineno}
\usepackage{bbm}
\usepackage{amsthm}
\usepackage{enumerate}
\usepackage{float}
\usepackage{stfloats}
\usepackage[numbers]{natbib}
\setcitestyle{numbers,square}
\usepackage[english]{babel}
\usepackage{cases}
\usepackage{array}
\usepackage{amsthm}
\usepackage{amsmath}
\usepackage{amssymb}

\theoremstyle{definition}

\theoremstyle{definition}

\theoremstyle{definition}

\theoremstyle{definition}

\theoremstyle{definition}

\usepackage{braket}
\usepackage{textgreek}
\usepackage{mathtools, cuted}
\usepackage{amsfonts}
\usepackage{algorithm}
\usepackage[noend]{algpseudocode}
\usepackage{hhline}
\usepackage{float}
\usepackage[normalem]{ulem} % For strikthrough text
\usepackage{booktabs}
\usepackage{tabularx}

\algnewcommand\algorithmicinput{\textbf{Input:}}
\algnewcommand\Input{\item[\algorithmicinput]}
\algnewcommand\algorithmicoutput{\textbf{Output:}}
\algnewcommand\Output{\item[\algorithmicoutput]}

\usepackage [english]{babel}
\usepackage [autostyle, english = american]{csquotes}
\MakeOuterQuote{"}
\usepackage{hhline}
\usepackage{enumitem}
\usepackage{float}

\usepackage{caption}

\usepackage{subcaption}
\usepackage{amssymb}
\usepackage{siunitx}
\usepackage{tabstackengine}
\newcommand\AddLabel[1]{%
  \refstepcounter{equation}% increment equation counter
  (\theequation)% print equation number
  \label{#1}% give the equation a \label
}
\newcolumntype{C}{>{\hfill$\displaystyle}c<{$\hfill}}
\newcolumntype{R}{>{\hfill$\displaystyle}r<{$}}
\newcolumntype{L}{>{$\displaystyle}l<{$\hfill}}

\newcolumntype{Z}{>{\hfill$\displaystyle}X<{$\hfill}}
\newcolumntype{E}{>{\hfill$\displaystyle}X<{$}}
\newcolumntype{K}{>{$\displaystyle}X<{$\hfill}}

\usepackage{mathtools}

\DeclarePairedDelimiter\floor{\lfloor}{\rfloor}

\newfloat{model}{tbp}{lom}
\floatname{model}{Model}

\newcounter{submodel}[model]
\newfloat{submodel}{tbp}{losm}
\floatname{submodel}{Model}

\newcommand{\BC}{{B$\cdot$C~}}

\newcommand{\integers}{\mathbb{Z}}

\newcommand{\Jobs}{\mathcal{J}}

\newcommand{\Machines}{\mathcal{M}}

\newcommand{\setupMatrix}{\mathbb{S}}

\newcommand{\jm}{{jm}}
\newcommand{\jb}{{jb}}

\renewcommand{\bm}{{bm}}

\newcommand{\ffp}{{ff'}}
\newcommand{\of}{{0f}}
\newcommand{\og}{{0g}}
\newcommand{\fg}{{fg}}
\newcommand{\gf}{{gf}}

\newcommand{\Batches}{\mathcal{B}}

\newcommand{\Families}{\mathcal{F}}

\newcommand{\machineState}{{\mathfrak{f}}}

\newcommand{\stime}{\tau}  % Processing time
\newcommand{\ptime}{p}  % Processing time
\newcommand{\weight}{\omega}  % Priority
\newcommand{\releaseTime}{r}  % Release time
\newcommand{\pp}[1]{\left(#1\right)}
\newcommand{\SizeOf}{\texttt{sizeOf}}
\newcommand{\StartOf}{\texttt{startOf}}
\newcommand{\EndOf}{\texttt{endOf}}
\newcommand{\PresenceOf}{\texttt{presenceOf}}
\newcommand{\Span}{\texttt{span}}
\newcommand{\EndBeforeStart}{\texttt{endBeforeStart}}

\newcommand{\StartBeforeStart}{\texttt{startBeforeStart}}
\newcommand{\NoOverlap}{\texttt{noOverlap}}
\newcommand{\Synchronize}{\texttt{synchronize}}

\newcommand{\Alternative}{\texttt{alternative}}
\newcommand{\StateFunction}{\texttt{state}}
\newcommand{\CumulFunction}{\texttt{cumul}}
\newcommand{\Permutation}{\text{Perm}}
\newcommand{\pulse}{\text{pulse}}
\newcommand{\AlwaysEqual}{\texttt{alwaysEqual}}
\newcommand{\AlwaysIn}{\texttt{alwaysIn}}

\usepackage{accents}
\newcommand\munderbar[1]{%
  \underaccent{\bar}{#1}}
\usepackage{nicematrix}
\definecolor{coregray}{RGB}{240,240,240}    % very light gray
\definecolor{batchblue}{RGB}{180,210,250}   % soft blue, mid-gray when printed
\definecolor{syncrose}{RGB}{240,160,160}    % deeper rose, darker in grayscale

\newcommand{\nt}[1]{{#1}}

\journal{Operations Research Perspectives}

\makeatletter
\def\ps@pprintTitle{%
 \let\@oddhead\@empty
 \let\@evenhead\@empty
 \def\@oddfoot{}%
 \let\@evenfoot\@oddfoot}
\makeatother

\begin{document}

\begin{frontmatter}

\title{Constraint Programming Model\nt{s} For Serial Batch Scheduling With Minimum Batch Size}

\author[1]{Jorge A. Huertas}
\ead{huertas.ja@gatech.edu}

\author[1]{Pascal Van Hentenryck}
\ead{pvh@gatech.edu}

% Include full affiliation details for all authors

\address[1]{Georgia Institute of Technology, Atlanta, Georgia, United States}

\begin{abstract}
In serial batch (s-batch) scheduling, jobs are grouped in batches and processed sequentially within their batch. This paper considers multiple parallel machines, nonidentical job weights and release times, and sequence-dependent setup times between batches of different families. Although s-batch has been widely studied in the literature, very few papers have taken into account a minimum batch size, typical in practical settings such as semiconductor manufacturing and the metal industry. The problem with this minimum batch size requirement has been mostly tackled with dynamic programming and meta-heuristics, and no article has ever used constraint programming (CP) to do so. This paper fills this gap by proposing, three CP models for s-batching with minimum batch size\nt{: (i) an \textit{Interval Assignment} model that computes and bounds the size of the batches using the presence literals of interval variables of the jobs. (ii) A \textit{Global} model that exclusively uses global constraints that track the size of the batches over time. (iii) And a \textit{Hybrid} model that combines the benefits of the extra global constraints with the efficiency of the sum-of-presences constraints to ensure the minimum batch sizes. }The computational experiments on standard cases compare the \nt{three} CP model\nt{s} with two existing mixed-integer programming (MIP) models from the literature. The results demonstrate the versatility of the proposed CP model\nt{s} to handle multiple variations of s-batching; and \nt{their} ability to produce, in large instances, better solutions than the MIP models faster.
\end{abstract}


\end{frontmatter}

%\linenumbers
\section{Introduction} \label{sec: introduction}

In the current and highly competitive landscape of the manufacturing industry, companies are under growing pressure to minimize production costs and reduce cycle times. \nt{One effective strategy to improve efficiency is to process similar tasks, called \textit{jobs}, together in groups known as \textit{batches}} \cite{Monch2011-Survey}. \nt{There are two main ways to process these batches. In \textit{parallel batching} (p-batch), all jobs in a batch are processed simultaneously \cite{Fowler2022-SurveyP-Batching}. In contrast, in \textit{serial batching} (s-batch), jobs in a batch are processed sequentially one after another \cite{PottsKovalyov2000-BatchSchedulingReview}. The benefits of p-batching are obvious since it saves time by processing multiple jobs at once. Similarly, s-batching is especially useful when grouping similar jobs can prevent repetitive machine setups, which are time-consuming and costly} \cite{Wahl2024}.

\nt{Serial batching appears in many industries, including }metal processing \cite{Gahm2022}, additive manufacturing (3D printing) \cite{Gahm2022, Uzunoglu2023}, paint \cite{Shen2012} and pharmaceutical production \cite{Awad2022}, chemical manufacturing \cite{Karimi1995}, and semiconductor manufacturing \cite{Monch2004, SMT2020-Paper}. In semiconductor manufacturing, \nt{for example, microchips are} built on silicon wafers through repeated steps \nt{such as} diffusion, photolithography, etching, ion implantation, and planarization \cite{Monch2013-FabBook, Chiang2012}. P-batching is commonly studied in the diffusion operations \cite{Monch2013-FabBook, Chiang2012}, while s-batching is more typical in photolithography \cite{Monch2011-Survey} and ion implantation \cite{SMT2020-Paper}.

The s-batch problem falls under what is known in the scheduling literature as a \textit{family scheduling model} \cite{PottsKovalyov2000-BatchSchedulingReview}. In this framework, each job belongs to a specific family, which typically represents a shared machine setup or product recipe. Jobs are processed one after another on each machine, and switching between families often requires setup times, for example, cleaning the machine or reconfiguring it with the right setup for the next job \cite{Wahl2024}. To reduce these costly transitions, batches are usually formed using jobs from the same family, minimizing unnecessary setup operations \cite{Monch2011-Survey}.

In the s-batch literature it is common to find \nt{ maximum batch sizes} due to physical capacities of the machines that process the batches \cite{Wahl2023, Wahl2024}. On the other hand, \nt{minimum batch sizes} accommodate practical situations where a minimum work load is necessary to justify the usage of the machine to process a batch \cite{Sung1997-bLB}. \nt{For example, in metal processing, laser cutting machines cut out metal parts from base metal slides which define families depending on material type and thickness. The slide maximum capacity is approximated by the areas of the minimum bounding rectangle of each job shape (defining the minimum batch size) and the area of the base metal slides (defining the maximum batch size) \cite{Gahm2022, Wahl2023}. In industrial 3D printing, batch capacity requirements are approximated by the volumes of the jobs’ enclosing cuboids and the maximum batch size is given by the volumes of the surrounding box (printing area times height) defined by the production technology \cite{Gahm2022, Wahl2024}. In the ion implantation area of the semiconductor manufacturing process, dopant ions are added into a silicon wafer by accelerating them through an electric field, ultimately altering the electrical properties of the wafers \cite{Winkler2017}. The ion sources (e.g., Germanium or Phosphor) come from implant gases, which have to be changed in every setup. The maximum batch size is given by the volumes of gas introduced \cite{SMT2020-Paper}, while the minimum batch size is given by the minimum number of runs required to justify the gas change \cite{SMT2020-DataSpecification}.}

\nt{Most of the studies in the s-batch literature consider maximum batch sizes \cite{Wahl2024}. In contrast,} only few studies have considered minimum batch sizes \cite{Sung1997-bLB, Mosheiov2008-bLB, Chretienne2011-bLB, Hazir2014-bLB, Shahvari2016-bLB, Shahvari2017-bLB, Castillo2015-bLB}. \nt{Furthermore,} none of them has ever used Constraint Programming (CP) to do so. This paper fills this gap by proposing \nt{three} CP models for s-batch scheduling with minimum batch size\nt{: The \textit{Interval Assignment}, \textit{Global}, and \textit{Hybrid} models.} The paper conducts thorough computational experiments that compare the proposed CP models with two existing mixed-integer programming (MIP) models from the literature that solve different variations of s-batch independently. These experiments demonstrate the versatility of the CP model\nt{s} to handle different variations of s-batch, and also \nt{their} ability to find high-quality solutions quickly.

The remainder of this paper is organized as follows. Section \ref{sec: lit review} presents a literature review on s-batch variations and studies that consider a minimum batch size. Section \ref{sec: contributions} clearly outlines the contributions of this paper. Section \ref{sec: problem description} formally presents the description of the problem addressed. Section \ref{sec: cp model} presents the \nt{Interval Assignment CP model. Section \ref{sec: global} presents the Global and Hybrid models.} Section \ref{sec: experiments} presents the computational experiments conducted and their results. Finally, Section \ref{sec: conclusion} presents the conclusions and outlines future lines of research.

\section{Literature review} \label{sec: lit review}

\begin{figure*}[t]
    \centering
    \begin{minipage}[t]{0.48\textwidth}
        \centering
        \begin{subfigure}[b]{0.45\linewidth}
            \centering
            \includegraphics[width=\textwidth]{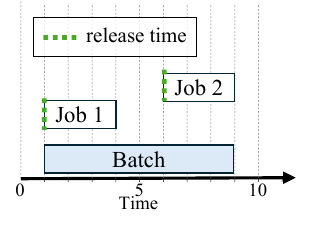}
            \caption{Preemptive}
            \label{fig: preemptive}
        \end{subfigure}
        \hfill
        \begin{subfigure}[b]{0.45\linewidth}
            \centering
            \includegraphics[width=\textwidth]{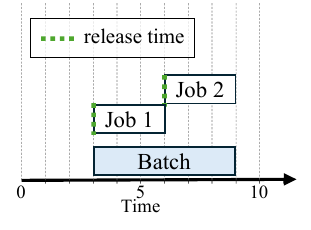}
            \caption{Non-preemptive}
            \label{fig: non-preemptive}
        \end{subfigure}
        \caption{Batch processing type}
        \label{fig: s-batch processing types}
    \end{minipage}
    \hfill
    \begin{minipage}[t]{0.48\textwidth}
        \centering
        \begin{subfigure}[b]{0.45\linewidth}
            \centering
            \includegraphics[width=\textwidth]{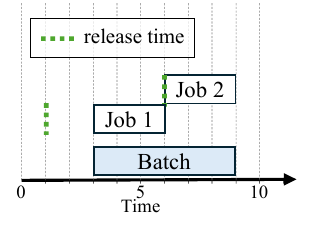}
            \caption{Flexible}
            \label{fig: flexible}
        \end{subfigure}
        \hfill
        \begin{subfigure}[b]{0.45\linewidth}
            \centering
            \includegraphics[width=\textwidth]{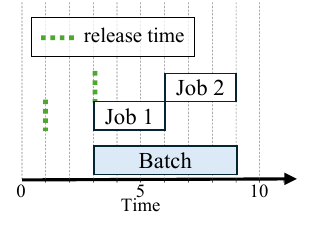}
            \caption{Complete}
            \label{fig: complete}
        \end{subfigure}
        \caption{Batch initiation type}
        \label{fig: s-batch initiation types}
    \end{minipage}
\end{figure*}

The scheduling literature distinguishes multiple s-batch variations \cite{Wahl2024}. Two variations exist depending on the time when jobs are considered completed: under \textit{item availability}, the jobs become completed as soon as their processing time is finished; instead, under \textit{batch availability}, jobs are considered completed when the entire batch has been processed \cite{PottsKovalyov2000-BatchSchedulingReview}. Figure \ref{fig: s-batch processing types} shows two types of variations depending on whether idle times are allowed to preempt the processing of jobs inside a batch. \textit{Preemptive} processing allows idle times inside a batch, and \textit{non-preemptive} forbids them \cite{Jordan1996}. Figure \ref{fig: s-batch initiation types} shows the two types of s-batch variations depending on the batch initiation. \textit{Flexible initiation} allows the batches to start before the release time of one (or more) of its jobs, while a \textit{complete initiation} forces all the jobs in the batch to be released before the batch start time. 

\nt{The typical s-batch variation in the photolithography and ion implant operations usually considers item availability, preemptive processing, and flexible initiation \cite{Monch2011-Survey}. For this reason, this variation is henceforward referred to as the IPF variation. On the other hand, the typical s-batch variation in the metal processing and pharmaceutical industry usually considers batch availability and complete initiation \cite{Gahm2022}. For this reason, this variation is henceforward referred to as the \BC variation. When using a utilization or cycle time objective (e.g., minimizing the makespan or the total weighted completion time (TWCT) \cite{Fowler2022-SurveyP-Batching}), the \BC variation results a non-preemptive batch processing. This happens because, under these types of objectives, the jobs are pulled to the start of the scheduling horizon as much as possible, squeezing out the idle spaces from the batches.}

Typically, the models proposed in different studies focus only on one specific variation of the problem and are unable to solve other variations. For example, \citet{Shahvari2017-bLB} proposed a MIP model \nt{for the IPF s-batch variation} that uses continuous variables for the completion times of the batches and the completion times of the jobs inside their batch; and binary variables to define the relative positioning of any pair of batches and any pair of jobs inside a batch. Because of this, their model is henceforward referred to as the \textit{Relative Positioning} (RP) MIP model. Minimal changes to the RP model allow it to consider batch availability. \nt{Nonetheless, due to its} lack of variables representing the start time of the batches or the start time of the jobs, \nt{it cannot handle the \BC s-batch variation.}

On the other hand, \citet{Gahm2022} proposed a MIP model \nt{for the \BC s-batch variation that assumes that all jobs are released at time 0 and they do not consider minimum batch sizes, just maximum. Furthermore, their way of modeling the batches is by predefining} the positions of the batches on the machines\nt{, i.e., by machine and by position on this machine}. Hence, this model defines binary variables to assign jobs to the $b$\textsuperscript{th} batch on each machine, scheduling empty batches at the end. Because of this, their model is henceforward referred to as the \textit{Positional Assignment} (PA) MIP model. Besides the binary assignment variables, the PA model also uses continues variables that capture the completion times of the batches; and additional variables are necessary to capture the batches' start times and consider non-identical release times. \nt{Furthermore, they proposed a constructive heuristic that is capable of solving instances with up to a thousand jobs, but relying on the simplification assumptions such as identical release times and not considering minimum batch size requirements. Additionally}, the PA model does not capture the order in which the jobs are processed inside the batches, making it unsuitable for the IPF variation, where knowing the end time of the jobs is necessary. \nt{To allow the RP and PA models solve other s-batch variations than their originally intended ones, significant changes are required in their formulations that completely modify their model structure, not only including (removing) additional (existing) variables and/or constraints, but also restructuring the existing ones.}

Existing s-batch reviews in the literature include the early one by \citet{PottsKovalyov2000-BatchSchedulingReview} in 2000, the one by \citet{Monch2011-Survey} in 2011, and the most recent by \citet{Wahl2024} in 2024. In the most recent one, only seven articles are known to consider a minimum batch size \cite{Sung1997-bLB, Mosheiov2008-bLB, Chretienne2011-bLB, Hazir2014-bLB, Shahvari2017-bLB, Shahvari2016-bLB, Castillo2015-bLB}. \citet{Sung1997-bLB, Mosheiov2008-bLB, Chretienne2011-bLB} and \citet{Hazir2014-bLB} focus on s-batching on a single machine and identical release times of the jobs. The extension to multiple machines and non-identical release times was addressed by \citet{Shahvari2017-bLB} in their RP model, and also by \citet{Shahvari2016-bLB} in a larger hybrid flowshop environment. The proposed approaches to solve s-batch with minimum  batch size include dynamic programming (DP) \cite{Sung1997-bLB, Chretienne2011-bLB, Hazir2014-bLB}, heuristic algorithms \cite{Sung1997-bLB}, rounding algorithms for the cases where only an upper or a lower bound on the batch size was considered, separately \cite{Mosheiov2008-bLB}, MIP models \cite{Shahvari2016-bLB, Shahvari2017-bLB}, Tabu Search (TS) algorithms \cite{Shahvari2017-bLB}, and Genetic Algorithms (GA) \cite{Castillo2015-bLB}. \nt{For example, the TS algorithm by \citet{Shahvari2017-bLB} solves instances with up to 27 jobs, taking up to 13,000 seconds in some cases, and obtaining deviations of up to 16\% with respect to the solution of the RP model found by CPLEX.} Nonetheless, \nt{to the best of our knowledge,} no article has used CP for s-batch with minimum batch size, as evidenced in the lack of references mentioning CP in the review by \citet{Wahl2024}.

\section{Contributions} \label{sec: contributions}

This paper proposes, for the first time, \nt{three} CP model\nt{s} for s-batch scheduling that considers a minimum batch size. Unlike the existing MIP models in the literature that can only handle specific variations of s-batch, the proposed CP model\nt{s} \nt{are} versatile and can address all the possible variations with minimal changes. \nt{The first model is an \textit{Interval Assignment} model that uses interval variables to assign and sequence jobs to batches on machines, enforcing the batch size requirements with non-global constraints at the assignment level. The second model is a \textit{Global} model that exclusively uses global constraints to ensure the minimum batch size requirements at every point in time. The third model is a \textit{Hybrid} model that combines the benefits of the previous two: global constraints that provide a global perspective to the solver and assignment-level constraints that boost performance.} The paper further explores the impact of symmetry-breaking constraints in the search process. The computational experiments compare the CP model\nt{s} with two existing models in the literature, mainly the RP \cite{Shahvari2017-bLB} and the PA \cite{Gahm2022} models, in their respective s-batch variations. The results demonstrate the versatility of the CP model\nt{s} to handle multiple variations of s-batch, as well as \nt{their} ability to produce, in larger instances, better solutions than these MIPs faster. \nt{For completeness, the MIP formulations used for comparison are provided in the appendix, as the main focus of this paper is the development and evaluation of the CP models.}

\section{Problem description} \label{sec: problem description}

Let $\Jobs$ be the set of jobs and $\Machines$ be the set of machines. Jobs are partitioned into families $\Families$ \nt{based on} their similarity. Let $f_j \in \Families$ be the family of job $j \in \Jobs$, and $\Jobs_f = \set{j \in \Jobs : f_j = f}$ be the subset of jobs that belong to family $f \in \Families$. Each job $j$ has a weight $\weight_j$, a release time $\releaseTime_j$, and a processing time $\ptime_j$. All the jobs can be scheduled on all the machines, and each machine can process only one job at a time. Consecutive jobs of the same family $\nt{f}$ are a \textit{serial batch} and it is necessary \nt{to have} a minimum \nt{and maximum} number \nt{$l_f$ and $u_f$} of consecutive jobs in the batch before processing another batch \nt{($0 < l_f \leq u_f$)}.  No setup is required between consecutive jobs of the same batch. However, let $\stime_\fg$ be the \textit{family setup time} when a batch of family $g \in \Families$ is immediately preceded by a batch of a different family $f \in \Families$, or $\stime_\og$ if there is no preceding batch. It is assumed that these setup times satisfy the triangular inequality, meaning that $\stime_\ffp \leq \stime_\fg + \stime_{gf'}$. 

The objective is to minimize the total weighted completion time (TWCT) of the jobs\nt{, which is the sum of the completion time of the jobs by their weight. The constraints include selecting the machine where each job is processed, ensuring that jobs processed in the same machine do not overlap while respecting the release times of the jobs, the family setup times between batches, and the minimum (and maximum) batch size requirements.}

\section{Constraint Programming model: Interval Assignment} \label{sec: cp model}

\nt{CP is a powerful method for solving combinatorial problems by defining constraints that a solution must satisfy. In CP, variables are assigned values from their domains, and the system works to ensure that all constraints are met. The search process explores the solution space in the form of a tree, where fixing the values of variables at each node branches out potential solutions. A key aspect of CP is the use of constraint propagation, which reduces the search space by pruning values from variable domains that cannot satisfy the constraints. Backtracking occurs when a branch leads to an invalid solution, allowing the search to revert to an earlier state and explore different variable assignments \cite{van2006handbook}.}

\setcounter{model}{1}
\begin{subequations}\label{eq:CP}
\begin{submodel}[thbp]
\caption{\nt{Core section}} \label{model:core}
\renewcommand{\arraystretch}{1.2}
% \resizebox{\textwidth}{!}{%
\begin{tabularx}{\linewidth}{@{}cL@{}L@{\hspace{0.5em}}KR@{}}
    \multicolumn{5}{@{}l}{\textbf{Variables and functions}:} \\
    \multicolumn{5}{@{}l}{\textbullet ~ Interval variables:} \\
    % x_j
    & x_j & \in \set{[s,s + \ptime_j): s \in \integers, \releaseTime_j \leq s}, & \forall ~j \in \Jobs; & \AddLabel{eq: cp def - interval job} \\
    % x_\jm
    & x_\jm & \in \set{[s,s + \ptime_j) : s \in \integers } \cup \set{\perp}, &  \forall ~ j \in \Jobs, m \in \Machines; & \AddLabel{eq: cp def - interval job on machine}\\
    \multicolumn{5}{@{}l}{\textbullet ~ Sequence variables:} \\
    % \varphi_m
    & \varphi_m & \multicolumn{2}{@{}L}{\in \Permutation(\set{x_\jm}_{j \in \Jobs}). \text{ Types: }\set{f_j}_{j \in \Jobs}, ~ \forall ~m \in \Machines;} & \AddLabel{eq: cp def - job sequence}\\
    % State Functions
    \multicolumn{5}{@{}l}{\textbullet ~ State functions:} \\
    % \machineState
    & \machineState_m &: \StateFunction & \forall ~ m \in \Machines; & \AddLabel{eq: cp def - state} \\
\end{tabularx}
\begin{tabularx}{\linewidth}{@{}cR@{}LK@{}R@{}}
    \hline
    \multicolumn{5}{@{}l}{\textbf{Formulation}:} \\
    & \multicolumn{3}{l}{\(\text{minimize~} \displaystyle\sum_{j \in \Jobs} \weight_j \cdot \EndOf(x_j)\)} & \AddLabel{eq: cp - obj function item} \\
    \multicolumn{5}{l}{subject to,}\\
    % Alternative machine for job
    & \multicolumn{2}{L}{\Alternative(x_j, \set{x_\jm}_{m \in \Machines}), \quad\quad} & \forall ~ j \in \Machines; & \AddLabel{eq: cp - job in one machine}\\
    % No overlapping jobs
    & \multicolumn{2}{L}{\NoOverlap(\varphi_m, \setupMatrix), }& \forall ~ m \in \Machines; & \AddLabel{eq: cp - no overlap jobs}\\
    % AlwaysEqual by jobs
    & \multicolumn{2}{L}{\AlwaysEqual(\machineState_m, x_\jm, f_j),} &  \forall ~ m \in \Machines, j \in \Jobs; & \AddLabel{eq: cp - state is job family} \\
\end{tabularx}
% }
\end{submodel}

\vfill

\begin{submodel}[thbp!]
\caption{\nt{Batching section}} \label{model:batching}
\renewcommand{\arraystretch}{1.2}
% \resizebox{\textwidth}{!}{%
\begin{tabularx}{\linewidth}{@{}cL@{}L@{}K@{}R@{}}
    \multicolumn{5}{@{}l}{\textbf{Additional variables and subsets}:} \\
    \multicolumn{5}{@{}l}{\textbullet ~ Interval variables:} \\
    % x_\jm^b
    & x_\jm^b & \multicolumn{2}{@{}L@{}}{\in \set{[s,s + \ptime_j) : s \in \integers} \cup \set{\perp},} & \\
    & & \multicolumn{2}{@{}E@{}}{\forall ~ j \in \Jobs, m \in \Machines, b \in \Batches_{f_j};} & \AddLabel{eq: cp def - interval job in batch on machine}\\
    % y_b
    & y_b & \multicolumn{2}{@{}L@{}}{\in \set{[s, e): s,e \in \integers, \stime_{0f^b} \leq s \leq e} \cup \set{ \perp }, ~ \forall ~ b \in \Batches;}  & \AddLabel{eq: cp def - interval batch}\\
    % y_\bm
    & y_\bm & \multicolumn{2}{@{}L@{}}{\in \set{[s, e): s,e \in \integers,  s \leq b} \cup \set{ \perp }, \forall ~ b \in \Batches, m \in \Machines;} & \AddLabel{eq: cp def - interval batch on machine}\\
    % Sets of interval variables
    \multicolumn{5}{@{}l@{}}{\text{\textbullet ~ Sets of intervals required to be present if }$x_\jm^b$\text{ is also present:}} \\
    & V_\jm^b & \multicolumn{2}{@{}L@{}}{= \set{y_b, y_\bm }, \quad \quad \forall ~ j \in \Jobs, m \in \Machines, b \in \Batches_{f_j};} & \AddLabel{eq: cp def - set of intervals}\\
    \multicolumn{5}{@{}l}{\textbullet ~ Sequence variables:} \\
    % \psi_m
    & \psi_m & \multicolumn{2}{@{}L@{}}{\in \Permutation(\set{y_\bm}_{b \in \Batches}). \text{ Types: }\set{f_j}_{j \in \Jobs}, \quad \forall ~ m \in \Machines;} & \AddLabel{eq: cp def - batch sequence}\\
\end{tabularx}
\begin{tabularx}{\linewidth}{@{}cR@{}L@{\hspace{0.5em}}K@{}R@{}}
    \hline
    \multicolumn{5}{@{}l}{\textbf{Additional constraints}:} \\
    % Alternative batch for job on machine
    & \multicolumn{2}{L}{\Alternative  (x_\jm, \set{x_\jm^b}_{b \in \Batches_{f_j}}),} & \forall ~ j \in \Jobs, m \in \Machines; & \AddLabel{eq: cp - alternative batch for job on machine}\\
    % Alternative machine for batch
    & \multicolumn{2}{L}{\Alternative (y_b, \set{y_\bm}_{m \in \Machines}),} & \forall ~ b \in \Batches; & \AddLabel{eq: cp - alternative machine for batch}\\
    % Implications
    &  \multicolumn{3}{L}{\PresenceOf(x_\jm^b) \Rightarrow \PresenceOf\pp{v},} & \\
    & & \multicolumn{2}{@{}R}{\forall ~ j \in \Jobs, m \in \Machines, b \in \Batches_{f_j}, v \in V_\jm^b;} & \AddLabel{eq: cp - implication}\\
    % Batch spans jobs on machine
    & \multicolumn{2}{L}{\Span (y_\bm, \set{x_\jm^b}_{j \in \Jobs_{f^b}}),} & \forall ~ b \in \Batches, m \in \Machines; & \AddLabel{eq: cp - span}\\
    % AlwaysEqual by batch
    & \multicolumn{2}{L}{\AlwaysEqual ( \machineState_m, y_\bm, f^b),} & \forall ~ m \in \Machines, b \in \Batches; & \AddLabel{eq: cp - state is batch family} \\
    % No overlapping batch
    & \multicolumn{2}{L}{\NoOverlap (\psi_m, \setupMatrix), } & \forall ~ m \in \Machines; & \AddLabel{eq: cp - no overlap batches}\\
\end{tabularx}
% }
\end{submodel}

\vfill

\begin{submodel}[thbp!]
\caption{\nt{Sizing section}} \label{model:batchsize}
\renewcommand{\arraystretch}{1.2}
% \resizebox{\textwidth}{!}{%
\begin{tabularx}{\linewidth}{@{}cR@{}L@{\hspace{0.5em}}K@{}R@{}}
    \multicolumn{5}{@{}l}{\textbf{Additional constraints}:} \\
    & \multicolumn{4}{L}{\forall ~ b \in \Batches:} \\
    & \sum_{j \in \Jobs_{f^b}}\sum_{m \in \Machines} \PresenceOf(x^b_\jm) & \geq l_{f^b} \cdot \PresenceOf(y_b) & & \AddLabel{eq: cp - presences sum lb}\\
    & \sum_{j \in \Jobs_{f^b}}\sum_{m \in \Machines} \PresenceOf(x^b_\jm) & \leq u_{f^b} \cdot \PresenceOf(y_b) & & \AddLabel{eq: cp - presences sum ub}
\end{tabularx}
% }
\end{submodel}

\nt{Scheduling is one of the most successful application areas of CP. It leverages \textit{interval variables} to represent tasks or operations over time. These interval variables encapsulate three key components: the start time, the end time, and the presence status of the interval (indicating whether the task is executed). The start time and the end time define the size of the interval. The presence attribute allows CP to model both mandatory and optional activities within a schedule. In particular, it plays a crucial role in handling optional tasks and alternative resource allocations. \textit{Interval sequence} variables allow to model sequences of tasks. CP also uses \textit{cumulative functions} to model the evolution of resource usage over time, ensuring that resource capacities are respected by aggregating the resource demands of all overlapping tasks. These cumulative functions are crucial in preventing over-utilization of resources like machines, personnel, or energy. The efficiency of the search process is further enhanced by global constraints that exploit the structure of the problem to propagate information and prune variable domains efficiently. Together, CP and its tools for constraint propagation, domain pruning, backtracking, and cumulative functions provide a structured and efficient approach for solving complex scheduling problems \cite{van2006handbook, Laborie2018}. The interested reader on an in-depth CP overview is referred to the well-known book by \citet{van2006handbook}, and the paper by \citet{Laborie2018}.}

\nt{Model \hyperref[model:core]{1} presents the mathematical formulation of the \textit{Interval Assignment} (IA) CP model for s-batching with minimum batch size for the \nt{IPF variation}. The syntax follows that of the IBM ILOG CPLEX CP Optimizer \cite{Laborie2018}, a widely used CP solver.
The model is organized into three sections:
(i) the \textit{Core} section, which sequences jobs on machines;
(ii) the \textit{Batching} section, which groups jobs into serial batches; and
(iii) the \textit{Sizing} section, which enforces minimum (and maximum) batch size requirements.
The name \textit{Interval Assignment} refers to the central modeling approach: jobs are represented by interval variables, and are assigned to batches through additional interval variables. These assignments capture both the temporal positioning and the grouping structure of the jobs within batches.

To illustrate how each section of the model contributes to constructing the solution, consider the simple example in Table~\ref{tab: example}, which involves 5 jobs belonging to 2 families, and a single machine. Assume the minimum batch sizes are $l_1 = 3$ for family 1 and $l_2 = 2$ for family 2. Also, let the initial setup times be $\tau_{0,1} = \tau_{0,2} = 1$, and the setup times between families be $\tau_{1,2} = \tau_{2,1} = 3$. The following sections describe each part of the model in detail and illustrate how the solution of this example evolves as each section is incrementally added.}

\begin{table}[t] 
\centering
\caption{\nt{Information of the illustrative example}}\label{tab: example}
\begin{tabular}{cccccc}
\toprule
\textbf{Job} & \textbf{Weight} & \textbf{Release Time} & \textbf{Processing Time} & \textbf{Family} \\
\midrule
1 & 1 & 1  & 2 & 1 \\
2 & 1 & 5  & 2 & 1 \\
3 & 1 & 6  & 2 & 2 \\
4 & 1 & 12 & 2 & 2 \\
5 & 1 & 11 & 2 & 1 \\
\bottomrule
\end{tabular}
\end{table}
% }

\subsection{Core Section}

\nt{The Core section }is a straight forward CP formulation that is commonly presented in many CP tutorials and official documentations \cite{IBM_CP_Optimizer_Sequence_Dependent_Setup_Times} to demonstrate how sequence-dependent setup times can be considered when sequencing non-overlapping jobs on machines. To do so, these setup times are packed in the matrix $\setupMatrix = \set{\stime_\fg}_{f,g\in \Families} \in \integers^{\Families \times \Families}$ that should satisfy the triangular inequality. \nt{This requirement avoids inconsistencies in setup durations, such as indirect transitions being shorter than direct ones, which would undermine the logic of the sequencing decisions.} \nt{The} Core \nt{section} schedules similar jobs \nt{consecutively }to avoid unnecessary setups. These consecutive jobs can be viewed as a serial batch. \nt{Thus, this Core section is itself a model for s-batching. Nonetheless, it} cannot guarantee the minimum batch size. To solve this problem, \nt{this paper proposes to extend the Core section with} the Batching \nt{and Sizing sections. The Batching} section \nt{includes additional variables and constraints to assign the jobs to non-overlapping batches and the Sizing section enforces the minimum batch size requirement.}

\nt{Model \ref{model:core} presents the Core section of the model.} Equation \eqref{eq: cp def - interval job} defines a variable of the form $[s,s + \ptime_j)$, i.e., it has a size of exactly $\ptime_j$ units of time, which represents job $j \in \Jobs$ and can only start after $\releaseTime_j$. Equation \eqref{eq: cp def - interval job on machine} defines an optional interval variable $x_\jm$ of size $\ptime_j$ that represents the option of job $j$ being processed on machine $m \in \Machines$. This interval is optional since it is allowed to take the value $\perp$, which indicates its absence from the solution. It is not necessary to indicate that variable $x_\jm$ can only start after $\releaseTime_j$ because\nt{, if selected to be present, }this variable is going to be synchronized with variable $x_j$, which already accounts for this. Equation \eqref{eq: cp def - job sequence} defines a sequence variable $\varphi_m$ of jobs on machine $m$, which is a permutation of the job intervals $\set{x_\jm}_{j \in \Jobs}$ on such machine, whose types are the associated job families. The last element of the core model is defined by equation \eqref{eq: cp def - state}, which defines a state variable $\machineState_m$ that represents the family being processed on machine $m$. This state function considers the transition times $\setupMatrix$ to change its values. 

Objective function \eqref{eq: cp - obj function item} minimizes the TWCT under item availability, tallying the completion of the jobs as soon as their processing time finishes.  Constraints \eqref{eq: cp - job in one machine} use the $\Alternative\nt{(v, V)}$ global constraint, which receives an interval variable $v$ and a set of optional intervals $V$. \nt{This constraint ensures that if} the interval $v$ is present in the solution, \nt{then}  exactly one interval from the set $V$ \nt{is selected} to be present in the solution as well, and synchronizes it with interval $v$. Hence, constraints \eqref{eq: cp - job in one machine} ensure that each job is processed on exactly one machine. Constraints \eqref{eq: cp - no overlap jobs} use the $\NoOverlap\nt{(\varphi, \setupMatrix)}$ global constraint, which receives an interval sequence variable $\varphi $ and a matrix with transition times $\setupMatrix$. This global constraint ensures non-overlapping intervals in the sequence defined by the permutation $\varphi$, with a minimum distance between them given by the transition times $\setupMatrix$. Hence, constraints \eqref{eq: cp - no overlap jobs} ensure that only one job is processed at a time on each machine, while respecting the family setup times. Constraints \eqref{eq: cp - state is job family} use the $\AlwaysEqual\nt{(h, v, a)}$ global constraint, which receives a state function $h$, an interval variable $v$, and an integer value $a$. This global constraint ensures that \nt{if $v$ is present in the solution, then the state function takes a constant value} $h\nt{(t)} = a$ \nt{at any point in time during interval} $v$\nt{, i.e., $t\in v$}. Hence, constraints \eqref{eq: cp - state is job family} ensure that the sate of each machine is the family of the job being processed.

\nt{The solution produced by the Core section for the example in Table~\ref{tab: example} is illustrated in Figure~\ref{fig:core solution}. Since there is only one machine, the Gantt chart represents its schedule. The figure shows how the job-on-machine intervals $x_\jm$ are assigned their optimal values, achieving a total weighted completion time (TWCT) of 55. This value corresponds to the sum of the completion times of all jobs, given that each job has a weight of 1.
Each job is shown as a box, with its color indicating its family: blue for family 1 and pink for family 2. Each job appears on a separate vertical level, and the green vertical lines mark their respective release times. The initial setup time is applied at time 0, allowing job 1 to start as soon as it is released. From there, jobs are scheduled sequentially without overlap. Setup times between families are respected—for example, between jobs 2 and 3, and again between jobs 4 and 5.
Although the minimum batch size for family 1 is $l_1 = 3$, the Core solution only places two consecutive jobs of that family (jobs 1 and 2) before switching to jobs of family 2. Job 5 from family 1 is scheduled last to minimize TWCT. Thus, the Core section effectively schedules non-overlapping jobs while respecting setup times between families. However, it cannot enforce minimum batch size requirements. For this reason, the Batching and Sizing sections are needed.}

\begin{figure}[t!]
    \centering
    \includegraphics[width=\linewidth]{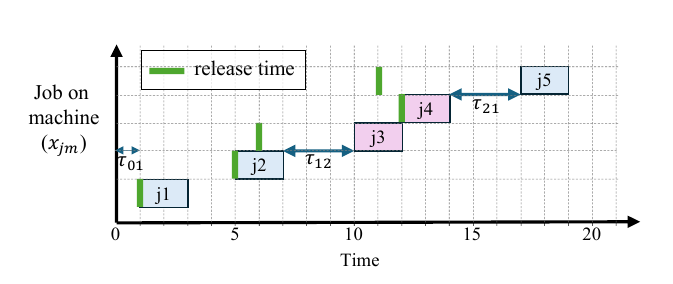}
    \caption{\nt{Solution of the Core section}}
    \label{fig:core solution}
\end{figure}

\subsection{Batching section}

\nt{The Batching and Sizing sections} use an ordered set of possible batches $\Batches = \set{1, 2, \ldots, N}$, where $N = \sum_{f \in \Families}N_f$ is the maximum number of possible batches that can be scheduled on a single machine, and $N_f = \floor{|\Jobs_f|/l_f}$ is the maximum number of possible batches needed to process jobs of family $f \in \Families$. This ordered set of possible batches can be further partitioned into mutually exclusive subsets by predefining the unique family $f^b \in \Families$ that each batch $b \in \Batches$ is allowed to process. Hence, $\Batches = \cup_{f \in \Families} \Batches_f$, where $\Batches_f = \set{b \in \Batches : f^b = f} \subset \Batches$ is the set of possible batches where jobs of family $f$ can be processed, and $|\Batches_f| = N_f$. In this way, the first $|\Batches_1|$ elements of $\Batches$ correspond to the possible batches where jobs of family $1$ can be processed; the next $|\Batches_2|$ elements to the possible batches for jobs of the family $2$, and so on.

\nt{Model \ref{model:batching} presents the Batching section} of the model. Equation \eqref{eq: cp def - interval job in batch on machine} defines another optional interval variable $x_\jm^b$ of size $\ptime_j$ that represents the option of job $j$ being processed on machine $m$ in batch $b \in \Batches_{f_j}$. \nt{It is not necessary to indicate that variable $x_\jm^b$ can only start after $\releaseTime_j$ because, if selected to be present, this variable is going to be synchronized with the present variable $x_\jm$, which is in turn synchronized with variable $x_j$ that already considers it. }Equation \eqref{eq: cp def - interval batch} defines an optional interval variable $y_b$ that represents batch $b \in \Batches$. To consider the initial setup time, this interval variable can only start after $\stime_{0f^b}$. It is optional because only batches with jobs assigned to it are present in the solution. Equation \eqref{eq: cp def - interval batch on machine} defines an optional interval variable $y_\bm$ that represents the option of sequencing batch $b$ on machine $m$. \nt{It is not necessary to specify that this interval can only start after $\stime_{0f^b}$ because, if selected to be present in the solution, it is going to be synchronized with variable $y_b$, which already considers it.
If interval variable $x_\jm^b$ is present in the solution it means that job $j$ is processed on machine $m \in \Machines$ in batch $b \in \Batches_{f_j}$. Thus, to ensure consistency, batch $b$ must be used and processed on machine $m$ as well. To ensure this, e}quation \eqref{eq: cp def - set of intervals} defines a set $V_\jm^b$ of interval variables that are required to be present in the solution if the interval variable $x_\jm^b$ is also present: \nt{the associated batch interval $y_b$ and its option on machine $y_\bm$.} Equation \eqref{eq: cp def - batch sequence} defines a sequence variable $\psi_m$ of batches on machine $m$, which is a permutation of the batch intervals $\set{y_\bm}_{b \in \Batches}$ on such machine, whose types are the associated batch families. 

Constraints \eqref{eq: cp - alternative batch for job on machine} ensure that if job $j$ is processed on machine $m$, it is processed in exactly one batch. Constraints \eqref{eq: cp - alternative machine for batch} schedule each batch on exactly one machine.  The machine that process each job (given by constraints \ref{eq: cp - job in one machine}), the batch where each job is processed (given by constraints \ref{eq: cp - alternative batch for job on machine}), and the machine that processes each batch (given by constraints \ref{eq: cp - alternative machine for batch}) are completely unrelated. Constraints \eqref{eq: cp - implication} solve this issue and guarantee that if a job $j$ is scheduled on a machine $m$ in batch $b$, i.e., $x^b_\jm$ is present, then the \nt{two} related intervals in $V^b_\jm \nt{= \set{y_b, y_\bm}}$ must also be present, linking them all together. Having ensured that the right intervals are present in the solution, constraints \eqref{eq: cp - span} ensure that the batch interval $y_\bm$ spans all the present job intervals $\set{x^b_\jm}_{j \in \Jobs_{f^b}}$ on the same machine in the same batch. These constraints use the $\Span\nt{(v, V)}$ global constraint, which receives an interval variable $v$ and a set of optional interval variables $V$. This global constraint ensures that the interval $v$ starts with the first present interval in $V$ and ends with the last present interval in $V$. \nt{In this way, constraints \eqref{eq: cp - span} capture the correct duration of the batches, based on the jobs assigned to them. }Constraints \eqref{eq: cp - state is batch family} add redundancy to the structure of the problem, which enhances computational performance as demonstrated by \citet{Huertas2024}. These constraints ensure that the state of the machines is the family of the batch being processed on them. Constraints \eqref{eq: cp - no overlap batches} include further redundancy and ensure that the batches being processed on the machines do not overlap, while respecting the setup family times. 

\nt{Figure~\ref{fig:batching solution} shows the solution of the illustrative example when combining the Core and Batching sections. The Batching section introduces new interval variables $x_\jm^b$, which indicate the specific batch $b$ in which each job $j$ is processed. These are displayed within the boxes as job–batch identifiers. The machine identifier is omitted since the example involves only a single machine.
In addition, new batch-level interval variables $y_\bm$ are introduced to represent the time span of each batch, covering the intervals of the jobs assigned to them. Batch 1 covers jobs 1 and 2, batch 3 covers only job 5, and batch 3 covers jobs job 3 and 4. Importantly, the Batching section does not alter the optimal schedule produced by the Core section; rather, it enriches the model with additional interval variables that explicitly capture the batch assignments.
While the Batching section defines which jobs belong to which batch, it does not enforce any batch size requirements. This responsibility is handled by the Sizing section, which is introduced next.}

\begin{figure}
    \centering
    \includegraphics[width=\linewidth]{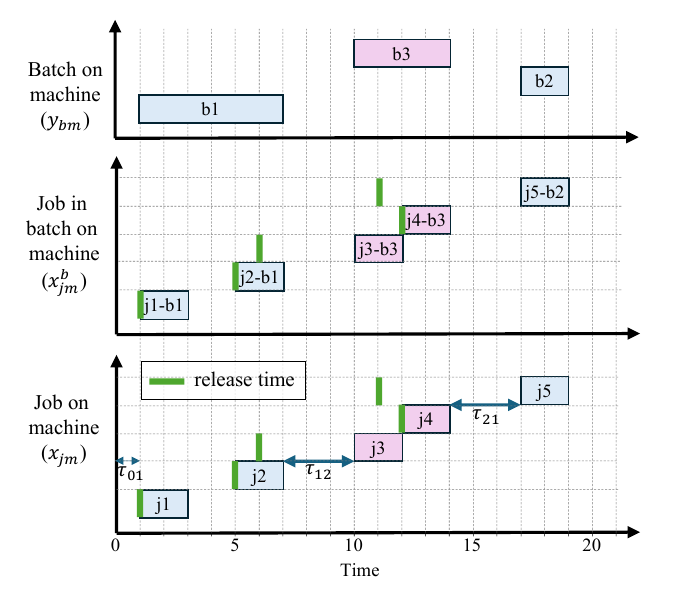}
    \caption{Solution of the Core and Batching sections}
    \label{fig:batching solution}
\end{figure}

\subsection{Sizing section}

\nt{The Batching section of the model assigns jobs to batches and captures the correct duration of the batches based on the jobs assigned to them. This Sizing section enforces the batch size requirements. Model \ref{model:batchsize} presents constraints \eqref{eq: cp - presences sum lb} and \eqref{eq: cp - presences sum ub}, which ensure the minimum and maximum batch size, respectively. The left-hand side of these constraints captures the size of a batch. The double summation holds because each batch is processed on at most one machine. The right-hand side of these constraints impose the minimum and maximum batch sizes, only if batch $b$ is in fact present in the solution.

Continuing with our example, the inclusion of the Sizing section in the model leads to a noticeable change in the solution, as the minimum batch size requirements are now enforced. Figure~\ref{fig:sizing solution} presents the complete solution obtained by the combined Core, Batching, and Sizing sections.
To satisfy the minimum batch size $l_1 = 3$ for family 1, the model must delay the processing of jobs 3 and 4 (from family 2) and schedule job 5 earlier. As a result, batch 1 now includes jobs 1, 2, and 5, which meets the minimum batch size requirement of  $l_1 = 3$ for family 1. Similarly, batch 3 includes jobs 3 and 4, satisfying the minimum batch size requirement of $l_2 = 2$ for family 2. These constraints are enforced by the Sizing section through the summation of presence variables, which ensures that only batches meeting the minimum size thresholds are allowed in the solution. Ensuring this constraint impacts the objective function, since the TWCT increases from 55 to 61.}

\begin{figure}[t!]
    \centering
    \includegraphics[width=\linewidth]{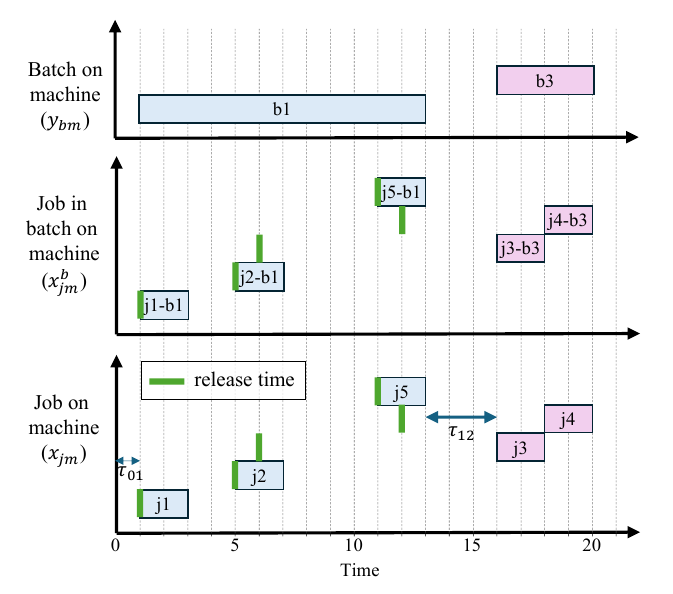}
    \caption{Solution of the Core, Batching and Sizing sections}
    \label{fig:sizing solution}
\end{figure}

\end{subequations}

\subsection{Handling problem variations}

\begin{subequations}\label{eq:IA  variations}
The following modifications are necessary to handle each problem variation, separately:
% \addtocounter{equation}{-1}

\begin{itemize}
    \item \textbf{Batch availability}: objective function \eqref{eq: cp - obj function item} should be replaced by objective \eqref{eq: cp - obj function batch ia}, which captures the completion time of the jobs as the completion time of their assigned batch.
        \nt{\begin{align}
            \text{minimize} \sum_{j \in \Jobs} \sum_{m \in \Machines} \sum_{b \in \Batches_{f_j}} \weight_j \cdot \PresenceOf(x^b_\jm) \cdot e_\jm^b, \label{eq: cp - obj function batch ia}
        \end{align}
        where $e_\jm^b$ is the end time of job $j$ if processed in batch $b \in \Batches_{f_j}$ on machine $m$. For the sake of space, equation \eqref{eq: cp - ia job end in batch} describes how to compute this value, which finds the maximum end time of all the jobs assigned to the batch in which job $j$ is processed.
        \begin{multline}
            e_\jm^b = \max_{k \in \Jobs_{f_j}} \left\{\EndOf(x_{km}^b) \cdot \PresenceOf(x_{km}^b)\right\}, \\
            \forall ~ j \in \Jobs, m \in \Machines, b \in \Batches_{f_j}.  \label{eq: cp - ia job end in batch}
        \end{multline}
        Once this new objective function is applied to the IA model, the new objective function of the example is 79, since jobs 1, 2, and 3 are only considered completed when the last job finishes, i.e., job 5 at time 13. And jobs 4 and 5 are considered completed when the last job finishes, at time 20.}
    \item \textbf{Non-preemptive processing}: constraints \eqref{eq: cp - non-preemption} should be included, which force the size of the interval $y_\bm$ to be the summation of the sizes of the present intervals $\set{x^b_\jm}_{j \in \Jobs_{f^b}}$. These constraints squeeze out of the batches any possible idle times.
        \begin{multline}
            \SizeOf(y_\bm) = \sum_{j \in \Jobs_{f^b}} \SizeOf(x^b_\jm),\\ \forall ~ b \in \Batches, m \in \Machines; \label{eq: cp - non-preemption}
        \end{multline}
    \nt{Figure \ref{fig:non-preemptive solution} presents the solution of the IA model when individually applying the non-preemptive constraints \ref{eq: cp - non-preemption}. To squeeze out the idle times inside batch 1, the processing of jobs 1 and 2 have to be delayed to synchronize the end of one job right before the start of the next one. These constraints cause that the new TWCT is 71.}
    \begin{figure}
        \centering
        \includegraphics[width=\linewidth]{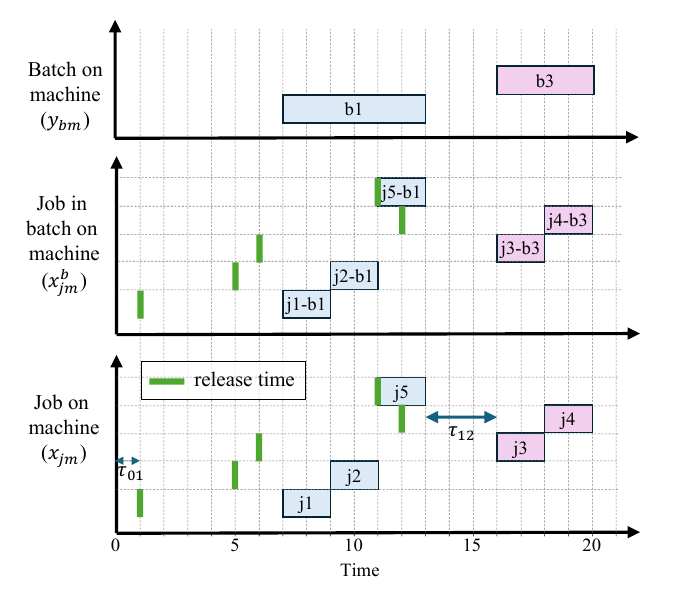}
        \caption{Solution with non-preemptive batch processing}
        \label{fig:non-preemptive solution}
    \end{figure}
    \item \textbf{Complete initiation}: 
    \nt{constraints \ref{eq: cp - complete initiation ia} should be included, which guarantee that a batch only starts on or after the release times of the jobs assigned to it.
    \begin{multline}
            \StartOf(y_\bm) \geq \releaseTime_j \cdot \PresenceOf(x_\jm^b), \\ \forall ~ j \in \Jobs, m \in \Machines, b \in \Batches_{f_j}. \label{eq: cp - complete initiation ia}
        \end{multline}
    Figure \ref{fig:complete solution} presents the solution of the IA model when individually applying constraints \ref{eq: cp - complete initiation ia}. To further ensure the complete batch initiation, jobs 1 and two have to be further delayed to start only after job 3 is released. These constraints cause that the new TWCT is 91.}
    \begin{figure}
        \centering
        \includegraphics[width=\linewidth]{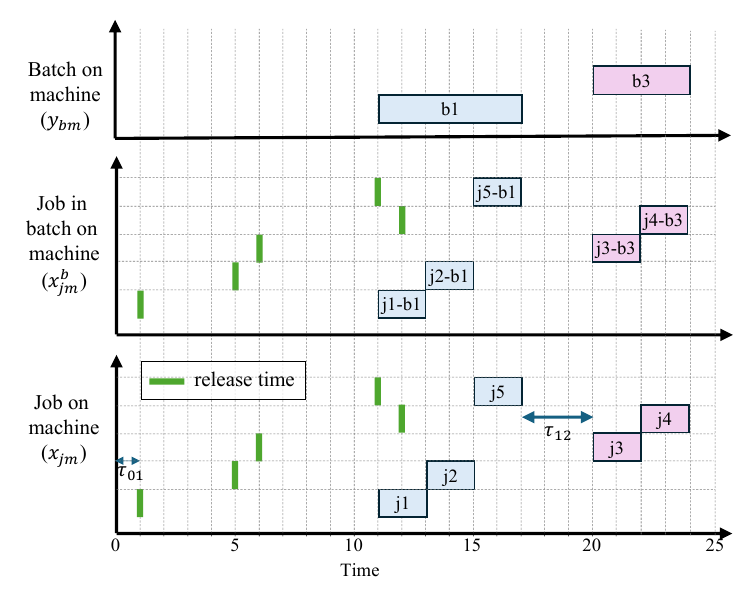}
        \caption{Solution with complete batch initiation}
        \label{fig:complete solution}
    \end{figure}
\end{itemize}

\end{subequations}

\subsection{Symmetry-breaking constraints}

\begin{subequations}\label{eq: symmetry-breaking constraints}
Symmetries in the search space could delay the CP engine to prove optimality. Hence, breaking these symmetries could potentially benefit the search process. Since the \textit{number} of the batch is irrelevant for the solution, the following symmetry-breaking SB constraints can be included in the CP model for any of the variants of s-batching:
\begin{multline}
    \PresenceOf(y_b) \leq  \PresenceOf(y_{b-1}),\\
    \forall ~ f \in \Families, b \in \Batches_f \setminus \min \Batches_f; \label{eq: cp - sb implication}
\end{multline}
\vspace{-3em}
\begin{multline}
    \StartBeforeStart(y_{b-1}, y_b), \\
    \forall ~ f \in \Families, b \in \Batches_f \setminus \min \Batches_f; \label{eq: cp - sb batch start}
\end{multline}
\vspace{-3em}
\begin{multline}
    \EndBeforeStart(y_\bm, y_{km}),\\
    \forall ~ m \in \Machines, f \in \Families, b,k \in \Batches_f ~|~  b < k. \label{eq: cp - sb machine batches}
\end{multline}

Constraints \eqref{eq: cp - sb implication} guarantee that non-used batches of each family are the last ones. Constraints \eqref{eq: cp - sb batch start} guarantee that the start times of the used batches of each family form a non-decreasing sequence. Constraints \eqref{eq: cp - sb machine batches} take these two concepts further and ensure that if any two batches of the same family are sequenced on the same machine, the batch with smaller \textit{number} is scheduled first.

\subsubsection{Tighter symmetry-breaking constraints}

In the \BC variation all the jobs in the same batch (which have already been released when the batch starts) are considered completed at the end of the last processed job. Therefore, the order in which these jobs are processed within the batch is not relevant for the objective function. Hence, besides the previous SB constraints, in this problem variation it is possible to enforce an additional tighter SB (SBT) constraint that forces an arbitrary order of the jobs inside a batch, e.g., by release time, as in constraints \eqref{eq: cp - sb order}.
\begin{multline}
    \EndBeforeStart(x_{im}^b, x^b_\jm), \\ \forall ~ m \in \Machines, b \in \Batches, i,j \in \Jobs_{f^b} ~| ~ \releaseTime_i \leq \releaseTime_j. \label{eq: cp - sb order}
\end{multline}

\end{subequations}

\section{Global and hybrid models} \label{sec: global}
\begin{subequations}\label{eq: model global model}

\nt{The global constraints in the Core and Batching sections of the IA model \hyperref[model:core]{1} rely on specialized algorithms that exploit the problem’s structure and provide the solver with a global perspective. This enables more effective constraint propagation and domain pruning. In contrast, the Sizing section of Model~\ref{model:batchsize} uses non-global constraints that enforce batch size requirements by summing the binary presence indicators of jobs within each batch and applying bounds, which do not leverage the advantages of the global reasoning.

To address this limitation, this section introduces the \textit{Global} model, which enforces the batch size requirements exclusively through global constraints. These constraints introduce additional structure and redundancy to the problem, which can improve computational performance in certain s-batching variations. However, the G model relies on cumulative functions that track the batch sizes continuously over time. This approach might be unnecessary, as the batch size only needs to be satisfied at the level of job assignment.

To balance expressiveness and efficiency, this section also introduces the \textit{Hybrid} model. This formulation combines the global constraints from the G model with the more direct constraints used in Model~\ref{model:batchsize}, which enforce batch size requirements without tracking them over time. The result is a model that retains the benefits of global propagation while avoiding unnecessary overhead.}

\subsection{Global model}

\nt{The \textit{Global} (G) CP model presented in this section enforces the size of the batches exclusively using global constraints. To do so, it uses the $\AlwaysIn(n,v,l,u)$ global constraints which receives a cumulative function $n$, an interval variable $v$, and integer values $l\leq u$. This global constraint guarantees that if $v$ is present in the solution, then the cumulative function $n$ is within the bounds $l$ and $u$ at any time during $v$, i.e., $l \leq n(t) \leq u ~\forall ~ t \in v$.

In this way, the cumulative function provided to the $\AlwaysIn$ constraint should capture the size of each batch so that the constraint can enforce the batch size requirements. However, cumulative functions in CP Optimizer only accumulate values over \textit{overlapping} intervals, whereas jobs within a serial batch are processed sequentially and do not overlap in time. To address this issue, the G model introduces additional \textit{virtual} interval variables that represent the full duration of the batch in which each job is processed. These virtual intervals are synchronized across all jobs in the same batch, ensuring they start and end at the same time. As a result, although the actual job intervals do not overlap, their corresponding virtual intervals do. This approach allows the cumulative function to correctly reflect the number of jobs in the batch at any moment during its execution.}

\nt{The G model replaces the Sizing section from Model \ref{model:batchsize} with the Global sizing section presented in Model \ref{model:batchsizeglobal}.}
Equation \eqref{eq: cp def - interval batched job} defines a \nt{virtual} interval variable $z_j$ for each job $j$, representing the full duration of the batch in which it is processed.
To model this duration correctly, equation \eqref{eq: cp def - interval job in batch} defines an optional \nt{virtual} interval $z_\jb$ for each possible batch $b \in \Batches_{f_j}$, indicating the batch-specific \nt{virtual} duration for job $j$.
These intervals are flexible in both start and end times, allowing them to span the entire batch, even before the job’s release time.
Equation \eqref{eq: cp def - set of intervals new} augments the set $V_\jm^b$ to include $z_\jb$, ensuring that whenever a job is scheduled in batch $b$ on machine $m$, its corresponding \nt{virtual} interval is also present.
Finally, equation \eqref{eq: cd def - cumul} defines a cumulative function $n_b$ for each batch $b$, which tallies the number of jobs assigned to it by pulsing 1 during their corresponding virtual intervals.

\setcounter{model}{3}
\begin{model}[t]
\caption{\nt{Global sizing section}} \label{model:batchsizeglobal}
\renewcommand{\arraystretch}{1.2}
% \resizebox{\textwidth}{!}{%
\begin{tabularx}{\linewidth}{@{}cL@{}L@{}K@{}R@{}}
    \multicolumn{5}{@{}l}{\textbf{Additional variables and functions}:} \\
    \multicolumn{5}{@{}l}{\textbullet ~ Interval variables:} \\
    % z_j
    & z_j & \in  \set{[s, e): s,e \in \integers, s \leq e}, & \forall ~ j \in \Jobs; & \AddLabel{eq: cp def - interval batched job}\\
    % z_\jb
    & z_\jb & \in  \set{[s, e): s,e \in \integers, s \leq e} \cup \set{ \perp }, & \forall ~ j \in \Jobs, b \in \Batches_{f_j}; & \AddLabel{eq: cp def - interval job in batch}\\
    % Sets of interval variables
    \multicolumn{5}{@{}l@{}}{\text{\textbullet ~ Sets of intervals required to be present if }$x_\jm^b$\text{ is also present:}} \\
    & V_\jm^b & \multicolumn{2}{@{}L@{}}{\gets V_\jm^b \cup \set{z_\jb}, \quad \quad \forall ~ j \in \Jobs, m \in \Machines, b \in \Batches_{f_j};} & \AddLabel{eq: cp def - set of intervals new}\\
    % Cumulative functions
    \multicolumn{5}{@{}l}{\textbullet ~ Cumulative functions:} \\
    % n_b
    & n_b &: \CumulFunction = \sum_{j \in \Jobs_{f^b}} \pulse \pp{z_\jb, 1}, & \forall ~ b \in \Batches. & \AddLabel{eq: cd  def - cumul}
\end{tabularx}
\begin{tabularx}{\linewidth}{@{}cR@{}L@{\hspace{0em}}K@{}R@{}}
    \hline
    \multicolumn{5}{@{}l}{\textbf{Additional constraints}:} \\
    % Alternative batch for batched job
    & \multicolumn{2}{L}{\Alternative (z_j, \set{z_\jb}_{b \in \Batches_{f_j}}),} & \forall ~ j \in \Jobs; & \AddLabel{eq: cp - alternative batch for batched job}\\
    % Synchronization
    & \multicolumn{2}{L}{\Synchronize(y_b, \set{z_\jb}_{j \in \Jobs_{f^b}}),} & \forall ~ b \in \Batches; & \AddLabel{eq: cp - synchronize}\\
    % AlwaysIn
    & \multicolumn{3}{L@{}}{\AlwaysIn(n_b, z_\jb, l_f, \nt{u_f}), \quad\quad~ \forall ~ f \in \Families, j \in \Jobs_f, b \in \Batches_f.} & \AddLabel{eq: cp - always in}\\
\end{tabularx}
\end{model}
\end{subequations}
\begin{figure}[t]
    \centering
    \includegraphics[width=\linewidth]{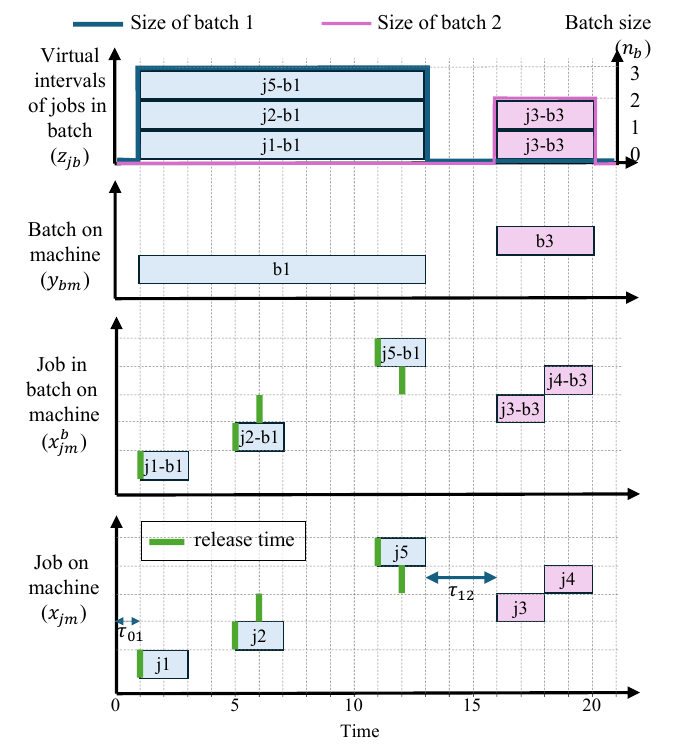}
    \caption{Solution of the Global model}
    \label{fig: global solution}
\end{figure}

\nt{Constraints \eqref{eq: cp - alternative batch for batched job} ensure that each job is associated with exactly one virtual interval $z_\jb$, and synchronize this selected interval with $z_j$.
Constraints \eqref{eq: cp - implication} then guarantee consistency by requiring $z_\jb$ to be present only when the job is actually assigned to batch $b$ on machine $m$, as dictated by the augmented set $V_\jm^b$.}
Constraints \eqref{eq: cp - synchronize} use the $\Synchronize\nt{(v,V)}$ global constraint, which receives an optional interval variable $v$ and a set of optional interval variables $V$. This global constraint aligns the start and end times of the present intervals in $V$ with the start and end times of the interval $v$, if present. Therefore, constraints \eqref{eq: cp - synchronize} ensure that the virtual intervals of all jobs in the same batch are synchronized with the batch interval $y_b$, \nt{effectively giving them the full duration of the batch. This alignment causes the virtual intervals to overlap, allowing the cumulative function $n_b$ to correctly represent the number of jobs in the batch over time.} This is an intuition derived from the CP model for p-batching presented by \citet{Huertas2024}. Lastly, constraints \eqref{eq: cp - always in} apply the $\AlwaysIn$ global constraint, enforcing that the batch size (as reflected by $n_b$) remains within the required lower and upper bounds throughout the interval.

\nt{Continuing with our example, Figure \ref{fig: global solution} presents the solution of the G model. The solution itself for the IPF variation doesn't necessarily change. The real change comes from the way the minimum batch size requirement is ensured. Figure \ref{fig: global solution} shows the new virtual variables $z_\jb$ created, which overlap for jobs in the same batch. This allows the cumulative functions $n_b$ to tally the correct size of the batches, and therefore bound them with the proper batch size requirements.}

\subsubsection{Handling problem variations}
\begin{subequations}\label{eq: global model}
\nt{To handle different s-batch variations with the G model, the same constraints \eqref{eq: cp - non-preemption} handle non-preemptive processing. In contrast, the following modifications are necessary to handle batch availability and complete initiation:} 
% \addtocounter{equation}{-1}

\begin{itemize}
    \item \textbf{Batch availability}: objective function \eqref{eq: cp - obj function batch ia} should be replaced by objective \eqref{eq: cp - obj function batch global}, which captures the completion time of the jobs as the completion time of their assigned batch using the virtual intervals.
        \begin{align}
            \text{minimize~} \sum_{j \in \Jobs} \weight_j \cdot \EndOf(z_j). \label{eq: cp - obj function batch global}
        \end{align}
    \item \textbf{Complete initiation}: instead of including constraints \eqref{eq: cp - complete initiation ia}, the domain definition of the virtual intervals $z_j$ in equation \eqref{eq: cp def - interval batched job} should be replaced by equation  \eqref{eq: cp def - interval batched job with preemption}. This new domain restricts the start time of the virtual intervals to start on or after the release time of the jobs, forcing the complete initiation.
        \begin{align}
            z_j &\in \set{[s,e): s,e \in \integers, \releaseTime_j \leq s \leq e}, & & \forall ~ j \in \Jobs. \label{eq: cp def - interval batched job with preemption}
        \end{align}
\end{itemize}
\end{subequations}

\subsection{Hybrid Model}

\nt{The \textit{Hybrid} (H) model combines the structural redundancy introduced by the global constraints of the G model with the simplicity of sum-of-presences constraints of the IA model, which enforce batch size requirements without tracking them over time. Specifically, the H model removes the cumulative functions~\eqref{eq: cd def - cumul} and the $\AlwaysIn$ constraints~\eqref{eq: cp - always in} from the G model, replacing them with the lower and upper bound constraints~\eqref{eq: cp - presences sum lb} and~\eqref{eq: cp - presences sum ub} from the IA model. These summation-based constraints ensure that batch sizes are respected while avoiding the overhead of time-dependent tracking.}

\nt{The H model is particularly effective for the BC variation, where jobs are considered completed only after the last job in their corresponding batch finishes. In this context, the order of jobs within a batch is irrelevant. The H model captures this behavior by shifting the objective to the virtual intervals, which abstract away the internal ordering of jobs, as evidenced by objective function \eqref{eq: cp - obj function batch global}. Nonetheless, the synchronization constraints~\eqref{eq: cp - synchronize} still link these orderless virtual intervals with the job sequencing, providing additional structure to the solver and a global perspective that enhances constraint propagation and domain pruning. Meanwhile, the sum-of-presences constraints enforce the batch size requirements without tracking them over time.}

\section{Computational experiments} \label{sec: experiments}

The computational experiments consider $1,170$ instances generated using \nt{a similar approach as the one used by \citet{Shahvari2017-bLB} in combination with the} process \nt{of} \citet{Huertas2024}, who create instances that gradually grow in size. The number of jobs in each instance can be one of four possible values: $|\Jobs| \in \set{15, 25, 50, 100}$. The number of families in each instance is $|\Families| \in \Phi_{|\Jobs|}$, where $\Phi_{15} = \set{2}$, $\Phi_{25} = \set{2,3}$, $\Phi_{50} = \set{3, 5}$, and $\Phi_{100} = \set{5, 7}$. The number of machines in each instance is $|\Machines| \in \Omega_{|\Jobs|}$, where $\Omega_{15} = \set{2}$, $\Omega_{25} = \set{2, 3}$, $\Omega_{50} = \set{3,4}$, and $\Omega_{100} = \set{4,5}$. All the processing times and job weights are generated as $\ptime_j, \weight_j \sim U([10])$, where $U([a])$ is the discrete uniform distribution over the set $[a] = \set{1, \ldots, a}$.

\nt{To generate integer family setup times that satisfy the triangular inequality, we first build a fully connected directed graph with $|\Families|$ nodes, where each arc is assigned a random weight uniformly drawn from the interval $[0, 1]$. Dijkstra's algorithm \cite{dijkstra1959note} is then executed from each node to compute the shortest-path distances to all other nodes. These minimum distances are scaled by a factor $S \in {20, 50, 100}$—as in \citet{Shahvari2017-bLB} and \citet{Schaller2000}—and rounded to the nearest integer. This rounding step can introduce violations of the triangular inequality, so we enforce the inequality whenever needed by checking which pairs of families have setup times that don't satisfy the inequality, and enforce the strict equality. This process might require multiple iterations and even restarts until a non-symmetric integer matrix $\setupMatrix$ is generated that satisfies the triangular inequality. This method does not yield setup times with an expected value of $(S+1)/2$; however, it does produce a realistic and consistent matrix of setup times suitable for modeling sequence-dependent setups with asymmetry and path-consistency.}

To generate the release times of the jobs, a lower bound of the overall makespan is computed as
{\small
\begin{equation}
    C_{\max} = \left\lceil \frac{\sum_{j \in \Jobs} \ptime_j + (|\Families|-1) \cdot \max_{f,g \in \Families} \stime_\fg + \max_{g \in \Families} \stime_\og}{|\Machines|} \right\rceil. \notag
\end{equation}
}
Hence, the release times are drawn from $U([C_{\max}])$. For each combination of the possible values for the number of jobs, families, machines, and setup times distributions, a total of 30 instances are generated, resulting in $1,170$ instances in total. 

To define the minimum batch sizes $l_f$, each instance was first solved using the Core \nt{Model \ref{model:core}}. From this solution, the minimum number of consecutive jobs of each family $f \in \Families$ scheduled on the machines was retrieved as $\munderbar{l}_f$. Then, the minimum batch size $l_f$ is defined as a random number between $\munderbar{l}_f + 1$ and $|\Jobs_f|$. \nt{This construction ensures that the specific solution found by the Core model is infeasible under the new batch size requirement,} thereby \nt{motivating} the usage of an extended model capable of explicitly enforcing these constraints. 

The experiments address the following two variations of s-batch, comparing the \nt{IA, G, and H} model\nt{s} with different MIP models in each variation:
\begin{itemize}
    \item \textbf{IPF}: item availability, preemptive processing, \& flexible initiation:
    \begin{itemize}
        \item RP: Relative Positioning MIP model by \citet{Shahvari2017-bLB} (see \ref{sec: RP model}).
    \end{itemize}

    \item \textbf{B$\cdot$C}: batch availability \& complete initiation:
    \begin{itemize}
        \item PA: Positional Assignment MIP model by \citet{Gahm2022} with additional variables and constraints to consider non-identical release times and minimum batch sizes (see \ref{sec: PA model}).
    \end{itemize}
\end{itemize}

All the models were implemented in \textsc{Python} 3.9.12. All the CP models were solved with IBM ILOG CP Optimizer \cite{Laborie2018} from \textsc{CPLEX} 22.1.1, using its \textsc{Python} interface \cite{ibm2023cpoptimizerpython}.
All the MIP models were solved with \textsc{Gurobi Optimizer} version 12.0.0 \cite{gurobi}.
\nt{A time limit of 1 hour was imposed to all the models. }
All the experiments were run on the PACE Phoenix cluster \cite{PACE}, using machines that run Red Hat Enterprise Linux Server release 7.9 (Maipo) with dual Intel\textregistered~ Xeon\textregistered~ Gold 6226 CPU @ 2.70GHz processors, with 24 cores, and 48 GB RAM, and parallelizing up to three experiments at the same time. Each run uses 8 cores and 16 GB RAM.
\begin{figure*}[t]
    \centering
    \begin{subfigure}[b]{0.85\linewidth}
        \centering
        \includegraphics[width=\textwidth]{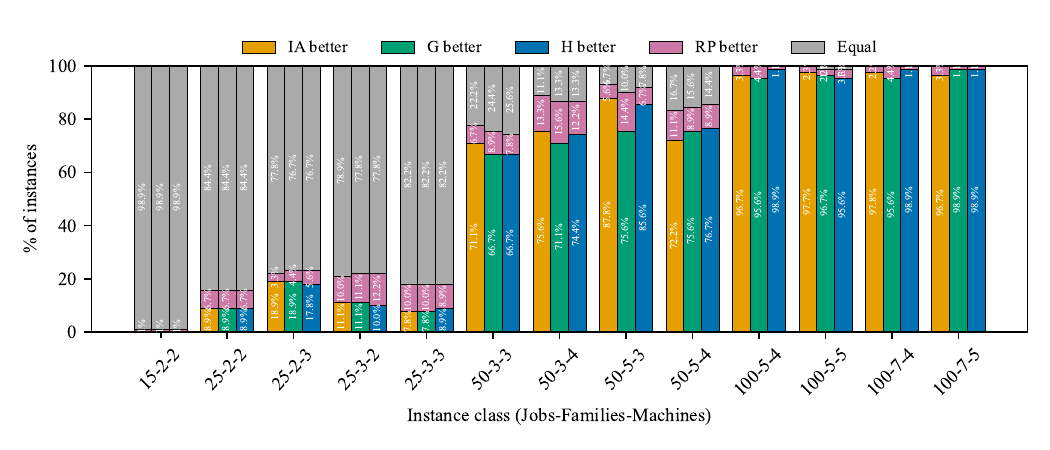}
        \vspace{-2.5em}
        \caption{IPF}
        \label{fig: better model IPF}
    \end{subfigure}
    \\
    \begin{subfigure}[b]{0.85\linewidth} % Adjust width to fit three subfigures
        \centering
        \includegraphics[width=\textwidth]{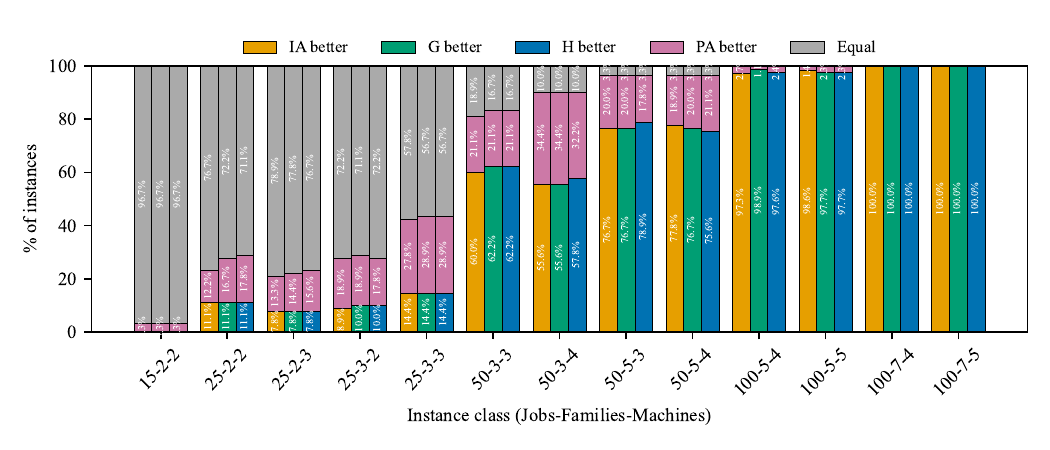}
        \vspace{-2.5em}
        \caption{B$\cdot$C}
        \label{fig: better model BC}
    \end{subfigure}
    \caption{Percentage (\%) of instances where each model is better or equal in each variation}
    \label{fig: better model}
\end{figure*}

Figure \ref{fig: better model} compares the results obtained with the proposed \nt{IA, G, and H} model\nt{s }against the MIP from the literature associated with each one of the two s-batch variations considered. These graphs group the 1,170 instances in 13 classes of 90 instances each according to their number of jobs, families, and machines. \nt{These graphs show three columns for each instance class in the horizontal axis. Each one of these columns is associated to one of the proposed CP models: the IA (in yellow), G (in green), and H (in blue) models, respectively. In this way, each column compares the column's CP model against the MIP model of the corresponding s-batch variation (in pink). Each cone of the columns displays three stacked bars. The first bar indicates the percentage of instances where the column's CP model obtained better solutions than the MIP model. The second bar indicates  the percentage of instances where the MIP model obtained better solutions than the column's CP model. Lastly, the last bar indicates the percentage of instances where both models obtained solutions with the same objective (in gray).} Both graphs reveal that \nt{in the small instances, the MIP and CP models find similar solutions, as evidenced by the dominance of the gray bars in the left-hand side of the graphs. Nonetheless, in instances with 50 and 100 jobs, the length of the MIP bars (pink) reduce in size. Thus, \textit{as the instance sizes grow, the CP models consistently outperform the MIP models}.}

\begin{figure*}[th]
    \centering
    \begin{subfigure}[b]{0.4\linewidth}
        \centering
        \includegraphics[width=\textwidth]{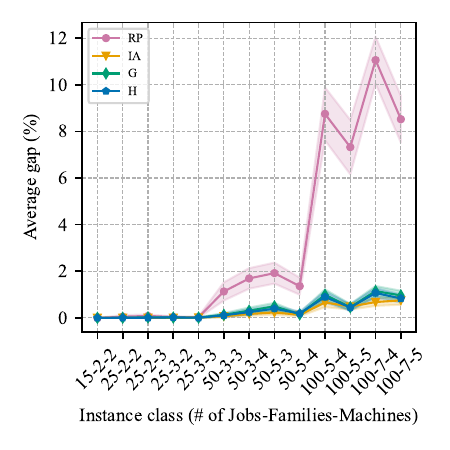}
        \vspace{-2.5em}
        \caption{IPF}
        \label{fig:gaps with MIP - IPF}
    \end{subfigure}
    \hspace{4em}
    \begin{subfigure}[b]{0.4\linewidth} % Adjust width to fit three subfigures
        \centering
        \includegraphics[width=\textwidth]{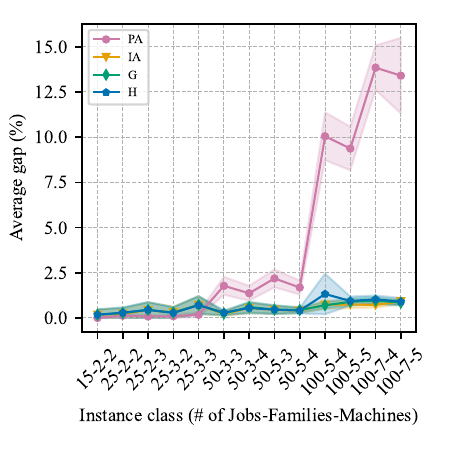}
        \vspace{-2.5em}
        \caption{B$\cdot$C}
        \label{fig:gaps with MIP - BC}
    \end{subfigure}
    \caption{Average relative gap (\%) of each model in each s-batch variation}
    \label{fig:gaps with MIP}
\end{figure*}

To asses the quality of the solution produced by each $model$ on each instance $i$, its \nt{\textit{relative}} gap is computed as $gap_{i,model} = |\text{TWCT}_{i,model} - \text{TWCT}_i^*|/|\text{TWCT}_{i,model}|$, where $\text{TWCT}_i^*$ is the minimum TWCT obtained for the instance $i$ across all the models that solved it (including those with SB and SBT constraints). Figure \ref{fig:gaps with MIP} shows the average \nt{relative} gaps of each one of the proposed CP models and the MIP model of the corresponding s-batch variation. \nt{Estimating the true relative gap would require infinite instances. Instead, the sample of 90 instances is used.} Thus, the region around each series represents the 95\% confidence interval of the avg. \nt{relative} gap. 

Figure \ref{fig:gaps with MIP - IPF} shows that in the IPF variation, on the instances with 15 and 25 jobs, the CP model\nt{s} and the RP model have the same gap of 0, meaning that they find solutions with the same TWCT, which corresponds to the results in Figure \ref{fig: better model}. On the other hand, Figure \ref{fig:gaps with MIP - BC} shows that in the \BC variation, on the instances with 15 and 25 jobs, the CP model\nt{s} have a slightly larger gap than the PA model. This is explained by the fact that the PA model does not consider in any way the decisions about ordering the jobs inside a batch whereas the CP models do. However, both graphs show that on the instances with 50 and 100 jobs, the CP model\nt{s} find better solutions than the MIP models. In fact, the solutions of the MIP models are up to 2\% worse than the solutions found by the CP models on the instances with 50 jobs; and up to 12\% worse on the instances with 100 jobs.

\nt{Figures \ref{fig:optimality}-\ref{fig: SB ctrs - BC} evaluate the impact of the symmetry-breaking constraints. Figures \ref{fig:optimality}-\ref{fig:time} focus on the small instances with 15 and 25 jobs. }Figure \ref{fig:optimality} shows the percentage of instances in which each model declared optimality within the time limit in each s-batch variation. Figure \ref{fig:time} shows the average solve time of each model in both s-batch variations. In both s-batch variations, no model is able to declare optimality on the large instances, hitting the time limit. However, on the small instances, the SB (and SBT) constraints consistently allow the CP models prove optimality in a larger percentage of instances than the CP models without these constraints. \nt{In the IPF variation, the IA+SB model is able to prove optimality in the largest percentage of instances, allowing it to terminate the search process faster, drastically dominating the RP model. In contrast, in the \BC variation, the best performing CP model is the H+SBT model, but it is still dominated by the PA model.}

\afterpage{
\begin{figure*}[thbp]
    \centering
    \begin{subfigure}[b]{0.4\linewidth}
        \centering
        \includegraphics[width=\textwidth]{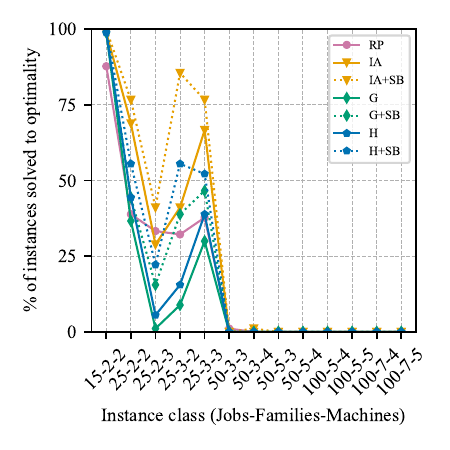}
        \vspace{-2.5em}
        \caption{IPF}
        \label{fig:optimality - IPF}
    \end{subfigure}
    \hspace{4em}
    \begin{subfigure}[b]{0.4\linewidth} % Adjust width to fit three subfigures
        \centering
        \includegraphics[width=\textwidth]{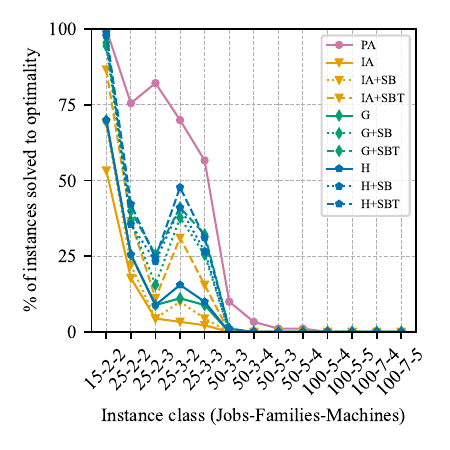}
        \vspace{-2.5em}
        \caption{B$\cdot$C}
        \label{fig:optimality - BC}
    \end{subfigure}
    \caption{Percentage (\%) of instances solved to optimality by each model in each s-batch variation}
    \label{fig:optimality}
\end{figure*}
\begin{figure*}[thbp]
    \centering
    \begin{subfigure}[b]{0.4\linewidth}
        \centering
        \includegraphics[width=\textwidth]{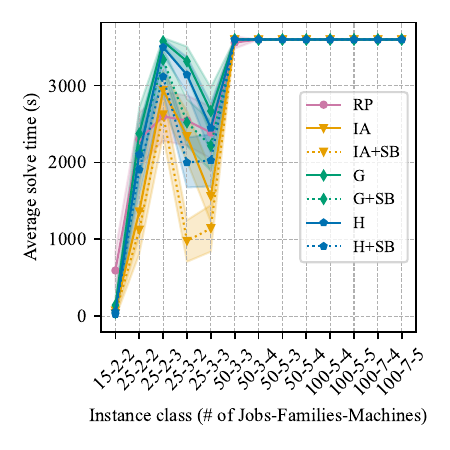}
        \vspace{-2.5em}
        \caption{IPF}
        \label{fig:time - IPF}
    \end{subfigure}
    \hspace{4em}
    \begin{subfigure}[b]{0.4\linewidth} % Adjust width to fit three subfigures
        \centering
        \includegraphics[width=\textwidth]{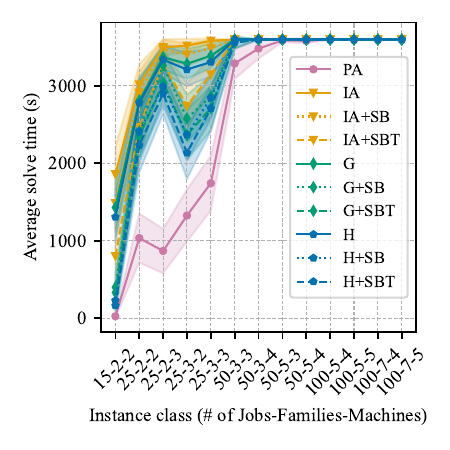}
        \vspace{-2.5em}
        \caption{B$\cdot$C}
        \label{fig:time - BC}
    \end{subfigure}
    \caption{Average solve time (s) of each model in each s-batch variation}
    \label{fig:time}
\end{figure*}
}

\afterpage{
\begin{figure*}[thbp]
    \centering
    \begin{subfigure}[b]{0.32\linewidth}
        \centering
        \includegraphics[width=\textwidth]{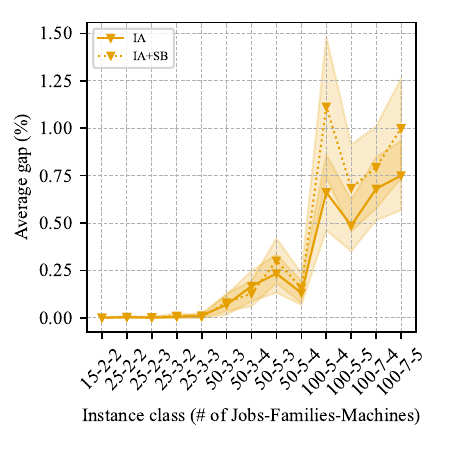}
        \caption{IA model}
        \label{SB ctrs IA - IPF}
    \end{subfigure}
    \hfill
    \begin{subfigure}[b]{0.32\linewidth} 
        \centering
        \includegraphics[width=\textwidth]{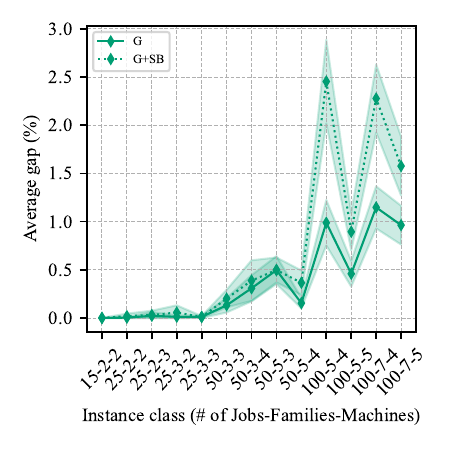}
        \caption{G model}
        \label{SB ctrs G - IPF}
    \end{subfigure}
    \hfill
    \begin{subfigure}[b]{0.32\linewidth}
        \centering
        \includegraphics[width=\textwidth]{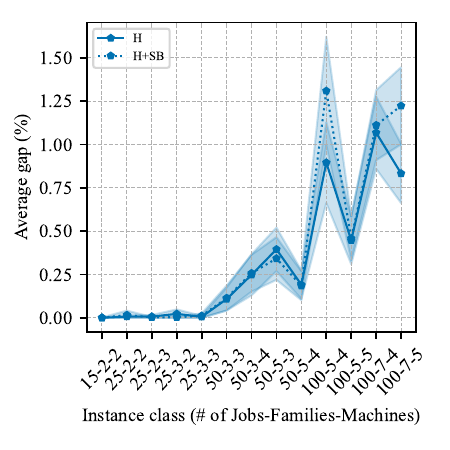}
        \caption{H model}
        \label{SB ctrs H - IPF}
    \end{subfigure}
    \caption{Impact of SB constraints on the relative gap of each model in the IPF s-batch variation}
    \label{fig: SB ctrs - IPF}
\end{figure*}
\begin{figure*}[thbp]
    \centering
    \begin{subfigure}[b]{0.32\linewidth}
        \centering
        \includegraphics[width=\textwidth]{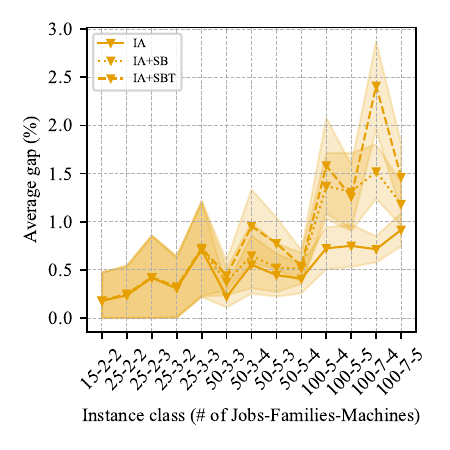}
        \caption{IA model}
        \label{SB ctrs IA - BC}
    \end{subfigure}
    \hfill
    \begin{subfigure}[b]{0.32\linewidth} 
        \centering
        \includegraphics[width=\textwidth]{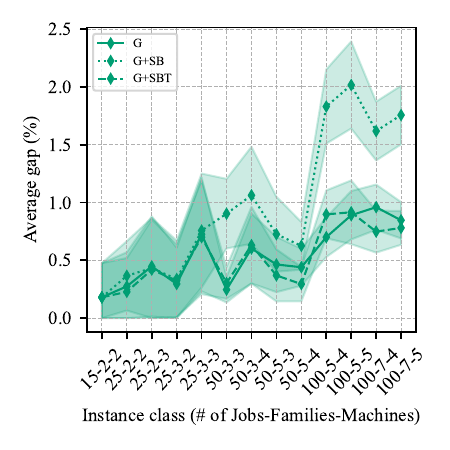}
        \caption{G model}
        \label{SB ctrs G - BC}
    \end{subfigure}
    \hfill
    \begin{subfigure}[b]{0.32\linewidth}
        \centering
        \includegraphics[width=\textwidth]{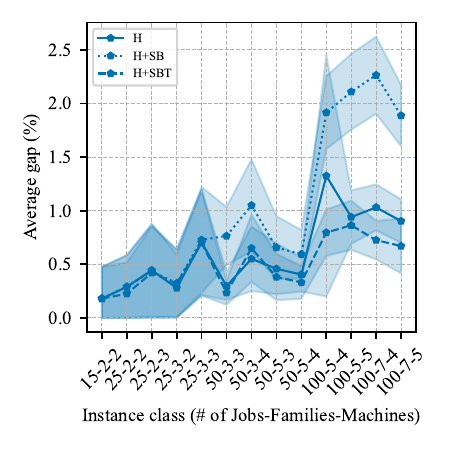}
        \caption{H model}
        \label{SB ctrs H - BC}
    \end{subfigure}
    \caption{Impact of SB and SBT constraints on the relative gap of each model in the \BC s-batch variation}
    \label{fig: SB ctrs - BC}
\end{figure*}
}

\nt{Since no model is capable of declaring optimality in the large instances with 50 and 100 jobs and all of them hit the time limit, Figures \ref{fig: SB ctrs - IPF} and \ref{fig: SB ctrs - BC} focus on evaluating the impact of the SB (and SBT) constraints using the relative gap of each CP model, independently. Figure \ref{fig: SB ctrs - IPF} demonstrates that in the IPF variation, on larger instances, the SB constraints consistently negatively impact the relative gap of each one of the proposed CP models. }\textit{Although the SB constraints help proving optimality in more of the small instances, allowing them to conclude the search process faster, these constraints become difficult to satisfy on the large instances under the same time limit, affecting their relative gap performance.}
\nt{Figure \ref{fig: SB ctrs - BC} shows that in the \BC variation, the IA model presents the same behavior. Nonetheless, the SBT constraints become helpful for the G and H models. In fact, the G+SBT model performs similar to the G model; whereas the H+SBT model becomes better than the H model. Two key observations can be derived from these results. The first one is that the extra redundancy provided by the global constraints in the G and H models prove to be beneficial when combined with the SBT constraints, which basically reduce the complexity of the decisions by forcing a specific order of jobs inside a batch. The second one is that this benefit is boosted in the H model because the batch size requirement is handled with the sum-of-presences constraints rather than tracking the batch size over time.}

\begin{figure*}[t]
    \centering
    \begin{subfigure}[b]{0.4\linewidth}
        \centering
        \includegraphics[width=\textwidth]{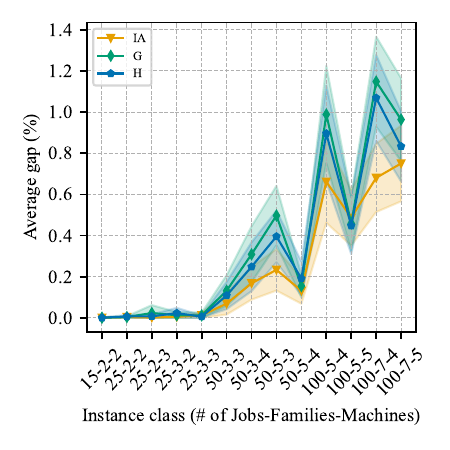}
        \vspace{-2.5em}
        \caption{IPF}
        \label{fig:gaps best - IPF}
    \end{subfigure}
    \hspace{4em}
    \begin{subfigure}[b]{0.4\linewidth} % Adjust width to fit three subfigures
        \centering
        \includegraphics[width=\textwidth]{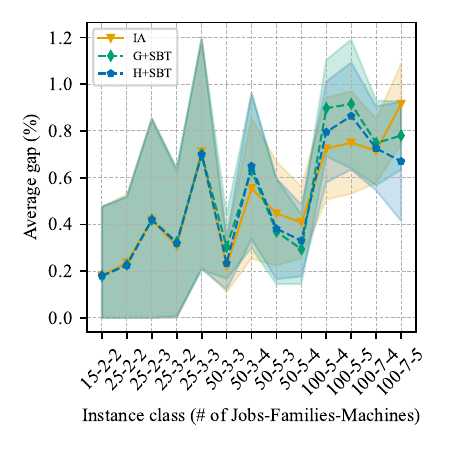}
        \vspace{-2.5em}
        \caption{B$\cdot$C}
        \label{fig:gaps best - BC}
    \end{subfigure}
    \caption{Avg. relative gap (\%) of the best CP model in each s-batch variation}
    \label{fig:gaps best}
\end{figure*}
\begin{figure*}[t]
    \centering
    \begin{subfigure}[b]{0.4\linewidth}
        \centering
        \includegraphics[width=\textwidth]{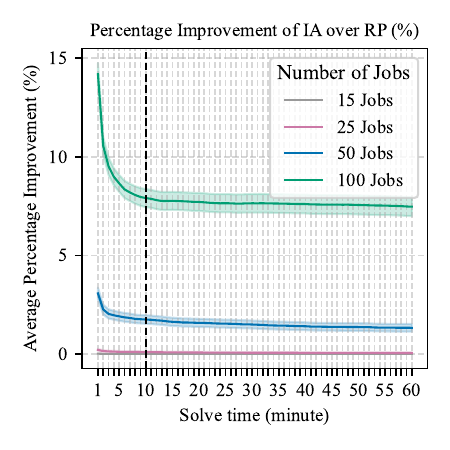}
        \vspace{-2.5em}
        \caption{IPF}
        \label{fig: minute improvement IPF}
    \end{subfigure}
    \hspace{4em}
    \begin{subfigure}[b]{0.4\linewidth} % Adjust width to fit three subfigures
        \centering
        \includegraphics[width=\textwidth]{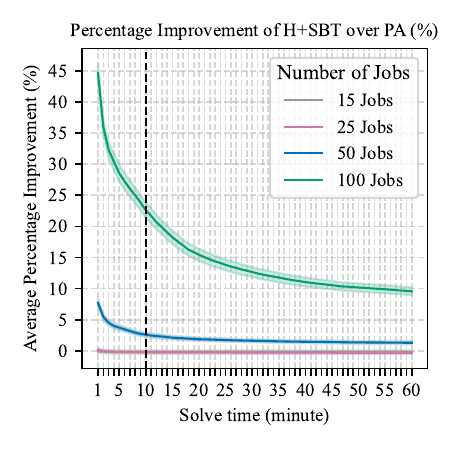}
        \vspace{-2.5em}
        \caption{B$\cdot$C}
        \label{fig: minute improvement BC}
    \end{subfigure}
    \caption{Average percentage (\%) of improvement of the best CP model in each s-batch variation over the corresponding MIP (with its 95\% CI) after every minute of solve time}
    \label{fig: minute improvement}
\end{figure*}

\nt{Figures \ref{fig: SB ctrs - IPF} and \ref{fig: SB ctrs - BC} show the performance with and without SB (and SBT) constraints of every CP model independently, which makes it difficult to determine which model is better. To remedy this, Figure \ref{fig:gaps best} presents the best of the IA, G, and H models in the same graph for every s-batch variation. For the IPF variation, the model with the best relative gap is the IA model, which clearly dominates the other models. In the \BC variation, it is not clear which model is better. Nonetheless, in the largest instance class, the best model is the H-SBT model.}

\nt{Although all the models reach the time limit on the larger instances, the benefits of CP go beyond merely finding better solutions within the time limit of 1 hour. They also find better solutions faster. }Figure \ref{fig: minute improvement} shows the average percentage of improvement (API) of the \nt{best-performing CP model of each s-batch variation over} the \nt{corresponding }MIP solution (API-CP-vs-MIP) and tracks how it evolves during the solve process, minute-by-minute. In these graphs, the 1,170 instances are grouped by the number of jobs in it, resulting in the four colored series for instances with 15, 25, 50, and 100 jobs. The shaded region around each series indicates the 95\% confidence interval (CI). To generate these graphs, for every instance $i$, the percentage of improvement of the CP solution over the MIP solution after $t$ minutes of solve time was computed as $(\text{TWCT}_{i,\text{MIP},t} -\text{TWCT}_{i,\text{CP},t}) / \text{TWCT}_{i,\text{MIP},t}$, where $\text{TWCT}_{i,model,t}$ is the best objective function value obtained by each model on each instance after $t$ minutes. This percentage is positive if the CP solution is better than the MIP solution, and negative in the opposite case. 

Figure \ref{fig: minute improvement} reveals that the API-CP-vs-MIP for instances with 15 and 25 jobs is near zero in both s-batch variations, reflecting that both models produce the same objective function values in most of these instances, as shown in Figure \ref{fig: better model}. Conversely, the API-CP-vs-MIP is significantly higher for instances with 50 and 100 jobs, where the \nt{best }CP model generally delivers better results. 
\nt{Since the scheduling systems in several areas of a wafer fab are expected to generate a Gantt-chart schedule every few minutes \cite{Ham2017-CP-p-batch-incompatible}, Figure \ref{fig: minute improvement} marks the API-CP-vs-MIP after 10 minutes of solve time.}
At 10 minutes,  the API-CP-vs-MIP for instances with 50 and 100 jobs is approximately 2\% and 7.5\%, respectively in the IPF variation; and 2\% and 24\%, respectively in the \BC variation. Hence, \textit{Figure \ref{fig: minute improvement} demonstrates the superiority of the \nt{best-performing }CP models over the MIP models in both s-batch variations, delivering better results in larger instances \nt{faster}}

\section{Concluding remarks} \label{sec: conclusion}

This paper proposed, for the first time, a \nt{set of }constraint programming (CP) model\nt{s} for serial batch scheduling, considering multiple parallel machines, non-identical job weights and release times, sequence-dependent setup times, and minimum batch sizes. A possible explanation to why no CP model has previously addressed s-batching with this minimum batch size can be the extra effort required in the modeling process to ensure this requirement. This becomes evident in the structure of the proposed CP model\nt{s}, which require \nt{a \textit{Batching} and a \textit{Sizing} section on top of the \textit{Core} section, each with }additional variables and constraints, just to ensure the minimum batch size requirement.

\nt{The proposed CP models are three. (i) An \textit{Interval Assignment} (IA) model that defines interval variables of jobs, and assigns them to batches with additional intervals. In this model, the minimum batch size requirement is enforced using bounding constraints over the sum of presence indicator of jobs in batches. (ii) A \textit{Global} (G) that replaces these constraints with additional global constraints that provide a global perspective of the problem's structure, allowing to better propagate constraints and prune variables domains. This G model tracks the size of the batches over time, which results impractical. (iii) The \textit{Hybrid} (H) model solves this issue by combining the strengths of the previous models, including the additional global constraints of the G model and using the efficient sizing constraints of the IA model.}

The proposed CP models can easily solve different variations of s-batch with minimal changes, including item and batch availability, preemptive and non-preemptive batch processing, and flexible and complete batch initiation. The computational experiments demonstrate this versatility by comparing the\nt{se} CP model\nt{s} \nt{against} two existing MIP models in the literature that are only capable of solving specific variations of s-batch independently. \nt{These experiments demonstrate that the best-performing model in the IPF variation is the IA model thanks to the efficient constraints that sum the presence indicators of the intervals in every batch. In the \BC variation, these constraints also allow the H-SBT model to have the best performance, in combination with the tighter symmetry breaking constraints and extra global perspective. }These experiments also showcase the ability of the CP model\nt{s} to find, in large instances, better solutions than these MIPs faster.

\nt{While this study offers a comprehensive comparison of different CP formulations across a wide range of instance sizes, future work could include a formal scalability and sensitivity analysis to assess memory usage and evaluate how variations in key parameters such as job weights, release times, and setup times affect performance.}

Current research includes \nt{improving the Global model for the IPF variation and considering family-machine dedications.} Future research includes embedding the serial and parallel batching CP models in larger contexts for wafer fab scheduling.

\section*{Acknowledgments}
This research was partly supported by the \href{https://www.ai4opt.org}{\textit{NSF AI Institute for Advances in Optimization}} (Award 2112533)

\appendix
\section{Relative Positioning model} \label{sec: RP model}
This section presents the Relative Positioning (RP) MIP model, inspired by \citet{Shahvari2017-bLB}. Their model only considered variables of completion times of jobs, which only allowed them to consider the problem variation with preemptive processing and flexible batch initiation. The key structure of the problem is provided by binary variables for every pair of batches and every pair of jobs inside the batches that indicate their relative position, which in turn are used to capture the completion times of the jobs, of the batches, and of the jobs inside the batches.

Let $x_\jb \in \set{0,1}$ be a binary variable that takes the value of 1 if job $j \in \Jobs$ is assigned to batch $b \in \Batches_{f_j}$, or 0 otherwise. Let $y_b \in \set{0,1}$ be a binary variable that takes the value of 1 if batch $b \in \Batches$ is used, or 0 if not. Let $y_\bm \in \set{0,1}$ be a binary variable that takes the value of 1 if the batch $ b \in \Batches$ is processed on machine $m \in \Machines$, or 0 otherwise. Let $z_{bij} \in \set{0,1}$ be a binary variable that takes the value of 1 if inside batch $b \in \Batches$ job $i \in \Jobs_{f^b}$ is sequenced before job $j \in \Jobs_{f^b}$ ($i<j$), or 0 otherwise. Let $w_{ab} \in \set{0,1}$ be a binary variable that takes the value of 1 if batch $a \in \Batches$ is before batch $b \in \Batches$ ($a<b$). Let $C_j$, $C^b$, and $C^b_j$, and  be three non-negative variables that represent the completion time of job $j \in \Jobs$, the completion time of batch $b \in \Batches$, and the completion time of job $j$ in batch $b \in \Batches_{f_j}$. Model \ref{model: RP} presents the mathematical formulation of the RP model for non-preemptive s-batch scheduling with item availability, preemptive processing, and flexible initiation.

\begin{model*}[!th]
\caption{Relative Positioning (RP) MIP model by \citet{Shahvari2017-bLB}} \label{model: RP}
\begin{mini!}[1]<b>
% Objective function
{}{\sum_{j \in \Jobs} \weight_j \cdot C_j \label{eq: RP - obj}}{}{}
% Each job in one batch
\addConstraint{\sum_{b \in \Batches_{f_j}}x_\jb}{= 1,}{\forall ~ j \in \Jobs;\label{eq: RP - each job in one batch}}
% Each batch in one machine
\addConstraint{\sum_{m \in \Machines} y_\bm}{= y_b,}{\forall ~ b \in \Batches;\label{eq: RP - each batch on one machine}}
% No job in batch if batch not used
\addConstraint{x_\jb}{\leq y_b,}{\forall ~ j \in \Jobs, b \in \Batches_{f_j};\label{eq: RP - no job if batch not used}}
% Minimum batch size requirement
\addConstraint{l_{f^b} \cdot y_b \leq \sum_{j \in \Jobs_{f^b}}x_\jb}{\leq |\Jobs_{f^b}| \cdot y_b, \quad}{\forall ~ b \in \Batches;\label{eq: RP - batch size requirements}}
% Batch sequencing 1
\addConstraint{C^b_j}{\geq C^a + \stime_{f^a,f^b} + \ptime_j - K \cdot [(1 - w_{ab}) + (1 - x_\jb)}{\notag}
\addConstraint{}{\quad\quad\quad\quad\quad\quad\quad\quad\quad\quad\quad + (1 - y_{am}) + (1 - y_\bm)],\quad}{\forall ~ a,b \in \Batches, j \in \Batches_{f^b}, m \in \Machines ~|~ a<b;\label{eq: RP - batch sequencing 1}}
% Batch sequencing 2
\addConstraint{C_j^a}{\geq C^b + \stime_{f^b,f^a} + \ptime_j - K \cdot [ w_{ab} + (1 - x_\jb) }{\notag}
\addConstraint{}{\quad\quad\quad\quad\quad\quad\quad\quad\quad\quad\quad + (1 - y_{am}) + (1 - y_\bm)],}{\forall ~ a,b \in \Batches, j \in \Batches_{f^a}, m \in \Machines ~|~ a<b;\label{eq: RP - batch sequencing 2}}
% Release times
\addConstraint{C^b_j}{\geq (\releaseTime_j + \ptime_j) \cdot y_b - K \cdot (1 - x_\jb),}{\forall ~ j \in \Jobs, b \in \Batches_{f_j}; \label{eq: RP - release times}}
% Initial setups
\addConstraint{C^b_j}{\geq (\stime_{0f_j} + \ptime_j) \cdot y_b - K \cdot (1 - x_\jb),}{\forall ~ j \in \Jobs, b \in \Batches_{f_j}; \label{eq: RP - initial setups}}
% Job sequencing 1
\addConstraint{C^b_j - C^b_i}{\geq \ptime_j \cdot y_b - K \cdot [(1 - z_{bij}) + (1-x_{ib}) + (1-x_\jb)], \quad}{\forall ~ b \in \Batches, i,j \in \Jobs_{f^b} ~|~ i<j;\label{eq: RP - job sequencing 1}}
% Job sequencing 2
\addConstraint{C^b_i - C^b_j}{\geq \ptime_i \cdot y_b - K \cdot [ z_{bij} + (1-x_{ib}) + (1-x_\jb)],\quad}{\forall ~ b \in \Batches, i,j \in \Jobs_{f^b} ~|~ i<j;\label{eq: RP - job sequencing 2}}
% Batch completion
\addConstraint{C_b}{\geq C^b_j - K(1-x_\jb),}{\forall ~ b \in \Batches, j \in \Jobs_{f^b};\label{eq: RP - batch completion}}
% Complete initiation
% \addConstraint{S^b}{\geq \releaseTime_j \cdot x_\jb,}{\forall ~ b \in \Batches, j \in \Jobs_{f^b}; \label{eq: RP - non-preemptiveness}}
% Non-preemption
% \addConstraint{C^b-S^b}{=\sum_{j \in \Jobs_{f^b}} \ptime_j \cdot x_\jb,}{\forall ~b \in \Batches; \label{eq: RP - non-preemptive}}
% Job Completion
\addConstraint{C_j}{\geq C^b_j - K(1-x_\jb),}{\forall ~ j \in \Jobs, b \in \Batches_{f_j};\label{eq: RP - item availability}}
% Variables domain
\addConstraint{C_j}{\geq 0, }{\forall ~ j \in \Jobs;\label{eq: RP - first domain}}
\addConstraint{C^b}{\geq 0, }{\forall ~ b \in \Batches;}
\addConstraint{C^b_j}{\geq 0, }{\forall ~ j \in \Jobs, b \in \Batches_{f_j};}
\addConstraint{x_\jb}{\in \set{0,1},}{\forall ~ j \in \Jobs, b \in \Batches_{f_j};}
\addConstraint{y_b}{\in \set{0,1},}{\forall ~ b \in \Batches;}
\addConstraint{y_\bm}{\in \set{0,1},}{\forall ~ b\in \Batches, m \in \Machines;}
\addConstraint{w_{ab}}{\in \set{0,1},}{\forall ~ a,b \in \Batches  ~|~ a<b;}
\addConstraint{z_{bij}}{\in \set{0,1},}{\forall ~ b \in \Batches, i,j \in \Jobs_{f^b}  ~|~ i<j. \label{eq: RP - last domain}}
\end{mini!}
\end{model*}

Objective function \eqref{eq: RP - obj} minimizes the TWCT. Constraints \eqref{eq: RP - each job in one batch} guarantee that each job is assigned to one batch. Constraints \eqref{eq: RP - each batch on one machine} ensure that if a batch is used, it is assigned assigned to exactly one machine. Constraints \eqref{eq: RP - no job if batch not used} prevent job assignments in batches that are not used. Constraints \eqref{eq: RP - batch size requirements} ensure the minimum batch size requirements. Constraints \eqref{eq: RP - batch sequencing 1}-\eqref{eq: RP - batch sequencing 2} are in charge of sequencing the batches. They ensure that the completion time of a job inside a batch is greater than the completion time of a previous batch, plus the family setup time required between the batches, plus the processing time of the job. These constraints use the big number $K$ to deactivate the constraint if the batches are not consecutive, or if the batches are not processed by the same machine, or if the job is not assigned to the batch. Constraints \eqref{eq: RP - release times} guarantee that the completion times of the jobs are after their release time plus their processing times. Constraints \eqref{eq: RP - initial setups} ensure that the initial setup times are satisfied. Constraints \eqref{eq: RP - job sequencing 1}-\eqref{eq: RP - job sequencing 2} are in charge of sequencing jobs inside their batch. They ensure that the difference between any pair of jobs inside a batch is at least their processing time. These constraints deactivate if the two jobs are not assigned to the same batch or if the relative positions of the jobs do not coincide. Constraints \eqref{eq: RP - batch completion} capture the completion time of the batches. Constraint \eqref{eq: RP - item availability} ensure item availability by tallying the overall completion time of the jobs as soon as their processing time has been completed. Finally, constraints \eqref{eq: RP - first domain}-\eqref{eq: RP - last domain} define the variables domain.

\section{Positional Assignment model} \label{sec: PA model}
This section presents the Positional Assignment (PA) MIP model, inspired by \citet{Gahm2022} for serial batching. Their original model assumes that all the jobs are available at time 0, and didn't consider a minimum batch size. The PA model in this section extends their formulation to consider the non-identical release times of the jobs, as well as the minimum batch size requirements. Instead of predefining the possible batches where jobs of each family can be grouped\nt{---as the IA, G, H, and RP models do}, this model assumes that these batches are sequentially scheduled on each machine according to their appearing order in the set $\Batches$. Hence, the family and the jobs allowed to be grouped in each batch are defined using binary assignment variables, scheduling empty batches after all the non-empty ones. This model does not capture the order in which the jobs are processed inside the batch. For this reason, it only works for s-batch variants with batch availability and complete batch initiation.

Let $x_\jm^b \in \set{0,1}$ be a binary variable that takes the value of 1 if job $j \in \Jobs$ is assigned to the $b$\textsuperscript{th} batch ($b \in \Batches$) on machine $m \in \Machines$, or 0 otherwise. Let $y^f_\bm \in \set{0,1}$ be a binary variable that takes the value of 1 if the $b$\textsuperscript{th} batch on machine $m$ processes jobs of family $f \in \Families$. Let $S_\bm, P_\bm$, and $C_\bm$ be three non-negative variables that represent the start, processing, and completion times of the $b$\textsuperscript{th} batch on machine $m$, respectively. Let $C_j$ be a non-negative variable that represents the completion time of job $j$. Model \ref{model: PA} presents the mathematical formulation of the PA model with batch availability and complete initiation.

\begin{model*}[!th]
\caption{Positional Assignment (PA) MIP model inspired by \citet{Gahm2022}} \label{model: PA}
\begin{mini!}[1]<b>
% Objective function
{}{\sum_{j \in \Jobs} \weight_j \cdot C_j \label{eq: PA - obj}}{}{}
% Each job on one machine and one batch
\addConstraint{\sum_{b \in \Batches}\sum_{m \in \Machines}x^b_\jm}{= 1, \quad}{\forall ~ j \in \Jobs;\label{eq: PA - job on 1 batch machine}}
% Each batch at most one family
\addConstraint{\sum_{f \in \Families}y^f_\bm}{\leq 1,}{\forall ~ b \in \Batches, m \in \Machines;\label{eq: PA - max 1 family per batch}}
% Jobs only in batches that attend the family
\addConstraint{x^b_\jm}{\leq y^f_\bm,}{\forall ~ f \in \Families, j \in \Jobs_f, b \in \Batches, m \in \Machines; \label{eq: PA - x < y}}
% Batch size requirements
\addConstraint{l_f \cdot y^f_\bm \leq \sum_{j \in \Jobs_f}x^b_\jm}{\leq |\Jobs_f| \cdot y^f_\bm, \quad}{\forall ~b \in \Batches, m \in \Machines, f \in \Families; \label{eq: PA - batch size requirements}}
% Empty batches at the end
\addConstraint{\sum_{f \in \Families}y^f_{b-1,m}}{\geq \sum_{f \in \Families}y_\bm^f, }{\forall ~ b \in \Batches, m \in \Machines ~|~ b > 1;\label{eq: PA - empty batches at the end}}
% % Processing time
\addConstraint{P_\bm}{\geq \sum_{j \in \Jobs}\ptime_j \cdot x^b_\jm,}{\forall ~ b \in \Batches, m \in \Machines; \label{eq: PA - processing time}}
% Initial setup
\addConstraint{S_{1,m}}{\geq \sum_{f \in \Families} \stime_\of \cdot y_{1,m}^f,}{\forall ~ m \in \Machines;\label{eq: PA - initial setup}}
% Intermediate setup times
\addConstraint{S_\bm}{\geq C_{b-1,m} + \stime_\gf - K \cdot [(1-y^g_{b-1,m}) + (1-y_\bm^f)],\quad\quad}{\forall ~ b \in \Batches, m \in \Machines, g,f\in \Families ~|~ b>1;\label{eq: PA - intermedium setup times}}
% Release times
\addConstraint{S_\bm}{\geq \releaseTime_j \cdot x_\jm^b,}{j \in \Jobs, b \in \Batches, m \in \Machines; \label{eq: PA - release times}}
% Batch completion times
\addConstraint{C_\bm}{\geq S_\bm + P_\bm,}{\forall ~ b \in \Batches, m \in \Machines; \label{eq: PA - batch completion times}}
% Completion times of the jobs
\addConstraint{C_j}{\geq C_\bm -K(1-x^b_\jm),}{\forall ~ j \in \Jobs, b \in \Batches, m \in \Machines; \label{eq: PA - job completion times}}
\addConstraint{x^b_\jm}{\in \set{0,1},}{\forall ~ j \in \Jobs, b \in \Batches, m \in \Machines;\label{eq: PA - first domain}}
\addConstraint{S_\bm, C_\bm, P_\bm}{\geq 0,}{\forall ~b \in \Batches, m \in \Machines;}
\addConstraint{C_j}{\geq 0,}{\forall ~ j \in \Jobs. \label{eq: PA - last domain}}
\end{mini!}
\end{model*}

Objective function \eqref{eq: PA - obj} minimizes the TWCT. Constraints \eqref{eq: PA - job on 1 batch machine} ensure that each job is assigned to exactly one batch on one machine. Constraints \eqref{eq: PA - max 1 family per batch} ensure that each batch groups jobs of up to one family. Constraints \eqref{eq: PA - x < y} ensure that jobs don't get assigned on batches that are not processing their family. Constraints \eqref{eq: PA - batch size requirements} ensure the minimum batch size requirements. Constraints \eqref{eq: PA - empty batches at the end} guarantee that empty batches are scheduled after nonempty batches on each machine. Constraints \eqref{eq: PA - processing time} capture the processing time of each batch. Constraints \eqref{eq: PA - initial setup} guarantee that the initial setup times are respected. Constraints \eqref{eq: PA - intermedium setup times} ensure that the intermediate family setup times are respected between consecutive batches. They use the big number $K$ to deactivate the constraint if the intermediate family setup time is . Constraints \eqref{eq: PA - release times} ensure the complete batch initiation. Constraints \eqref{eq: PA - batch completion times} capture the completion time of each batch. Constraints \eqref{eq: PA - job completion times} capture the completion time of each job using batch availability. Finally, constraints \eqref{eq: PA - first domain}-\eqref{eq: PA - last domain} define the variables' domain.

\bibliographystyle{elsarticle-num-names}
\bibliography{references}  % Ensure you have a mybib.bib file with your references.

\balance
\end{document}